\documentclass[aps,pre,reprint]{revtex4-1}

\usepackage{amstext}
\usepackage{amsmath}
\usepackage{amssymb}
\usepackage{amsfonts}
\usepackage{units}

\usepackage{graphicx}
\usepackage{color}
\usepackage[normalem]{ulem} 

\usepackage{microtype}
\usepackage[pdfborder={0 0 0}]{hyperref} 
\usepackage{mathtools}

\newcommand*{\diff}{\mathrm{d}}

\newcommand {\EHo} {E_\text{2D}}                 
\newcommand {\KHo} {K_\text{2D}}                 
\newcommand {\tEHo}{\tilde E_\text{2D}}
\newcommand {\tKHo}{\tilde K_\text{2D}}
\newcommand {\taus} {\tau^{\smash{(\text{s})}}}  
\newcommand {\tauc} {\tau^{\smash{(\text{c})}}}	 
\newcommand {\ttauc} {\tilde\tau^{\smash{(\text{c})}}}	 
\newcommand {\taub} {\tau^{\smash{(\text{b})}}}	 
\newcommand {\lambdab} {\lambda^{\smash{(\text{b})}}}	 

\newcommand {\qq}[1] {\quad \text{#1} \quad }   
\newcommand {\zweiervec}[2] {\begin{pmatrix}#1\\#2\end{pmatrix}}

\let\phi\varphi                     

\let\epsilon\varepsilon

\let\Psi\varPsi
\newcommand {\rA} {$\,\rightarrow\,$}
\newcommand {\brA} {$\,\boldsymbol\rightarrow\,$}

\definecolor{Red}{rgb}{0.9,0.0,0.1}
\definecolor{Lila}{rgb}{0.7,0.0,0.9}
\definecolor{Darkblue}{rgb}{0.22,0.33,0.64}
\definecolor{Darkgray}{rgb}{0.4,0.4,0.4}
\definecolor{Blue}{rgb}{0.1,0.0,0.9}

\begin{document}

\author{Sebastian Knoche} 
\affiliation{Department of Physics, Technische Universit\"{a}t Dortmund, 44221 Dortmund, Germany}

\author{Jan Kierfeld} 
\email{jan.kierfeld@tu-dortmund.de}
\affiliation{Department of Physics, Technische Universit\"{a}t Dortmund, 44221
  Dortmund, Germany}

\title{Elasticity of Interfacial Rafts of 
  Hard Particles with Soft Shells}

\begin{abstract}
  We study an elasticity model for compressed protein monolayers or particle
  rafts at a liquid interface. Based on the microscopic view of hard-core
  particles with soft shells, a bead--spring model is formulated and analyzed
  in terms of continuum elasticity theory. The theory can be applied, for
  example, to hydrophobin-coated air--water interfaces or, more generally, to
  liquid interfaces coated with an adsorbed monolayer of interacting hard-core
  particles. We derive constitutive relations for such particle rafts and
  describe the buckling of compressed planar liquid interfaces as well 
  as their apparent Poisson ratio. We also use
  the constitutive relations to obtain shape equations for 
  pendant or buoyant capsules
  attached to a capillary, and to compute deflated shapes of such capsules. A
  comparison with capsules obeying the usual Hookean elasticity (without hard
  cores) reveals that the hard cores trigger capsule wrinkling. Furthermore,
  it is shown that a shape analysis of deflated capsules with
  hard-core/soft-shell elasticity gives apparent elastic moduli which can be
  much higher than the original values if Hookean elasticity is
  assumed.
\end{abstract}

\maketitle

\section{Introduction}

Many soft matter systems exhibit elastic properties that go beyond the Hookean
linear elasticity. There are a number of prominent examples 
of  elastic materials with unique properties, which 
require tailored elasticity models for an accurate theoretical description. 
An early example is the Mooney-Rivlin law for large
deformations of incompressible, rubber-like materials \cite{Mooney1940}. To
describe the bending of lipid membranes, the Helfrich energy was 
introduced \cite{Helfrich1989},
and Skalak et al. and Evans developed a strain-energy function
for deformed red blood cell membranes \cite{Skalak1973,Evans1973a,Evans1973b}.
Only the combination of both types of elastic energy functionals 
  with a  Helfrich energy describing the lipid membrane and a  
 Skalak strain-energy function describing the spectrin skeleton
 is able to describe 
the experimentally observed sequences of red blood cell shapes 
successfully \cite{LimHW2002,Lim2008}.
The correct sequence of shapes cannot be reproduced using 
a simpler  effective model, for example, of the Helfrich type 
by choosing effective  bending moduli properly.

In general, 
indications whether simple elastic models such as 
a linear elasticity are sufficient  to describe the deformation of a
certain material deformation can be found when comparing experimental 
results to
theoretical predictions. If, for example, linear elasticity is assumed 
in the theoretical modeling but  
 comparison with experimental data shows that the resulting  elastic
moduli are apparently changing throughout the deformation,  this suggests
that a more advanced and detailed  elastic model should be used.

In this Article, we develop an elasticity model tailored for monolayers of
particles or molecules at a fluid interface under compression.
Since the pioneering work of Pickering \cite{Pickering1907}, it is known
that  adsorbed particles at a  liquid interface  act as surfactants and 
can stabilize droplets  in emulsions depending on the wettability 
properties \cite{Binks2002}.
Adsorbed particles  also stabilize foams \cite{Horozov2008}.
Colloidal particles adsorbing at a liquid droplet interface 
form colloidosomes \cite{Dinsmore2002}, where 
particles arrange to spherical colloidal crystals \cite{Bausch2003},
or ``armoured droplets'' if particles are jammed \cite{Subramaniam2005}.
A liquid interface  coated with adsorbed 
particles starts to  exhibit  elastic properties typical for a 
 two-dimensional solid or elastic membrane, such as resistance to 
shear, buckling, wrinkling and crumpling \cite{Vella2004,Zeng2006}.
This has also been reported for protein-coated bubbles \cite{Cox2007},
 droplets \cite{Stanimirova2011} or  vesicles \cite{Ratanabanangkoon2003}.

Hydrophobins are small proteins from fungal origin \cite{Wosten2001},
  which adsorb  strongly 
to the interface of an aqueous solution because of their amphipathic nature 
with a hydrophobic patch
on their surface \cite{Hakanpaa2004},
Layers of hydrophobin can be used very
efficiently to stabilize bubbles 
and foams \cite{Linder2005,Linder2009,Basheva2011},
 which makes them interesting as a model system for protein particle 
rafts at the air--water interface and for applications.

Particles coating liquid droplets  have applications 
in food industry \cite{Gibbs1999}.
Because of their strong amphipathic nature and biocompatibility,
hydrophobins  coating air--water interfaces 
 have various applications\cite{Hektor2005,Linder2005,Linder2009}
ranging from medical and technical coatings \cite{Lumsdon2005}
to the production of
   protein glue and cosmetics, or in the food
 industry in the stabilization 
of  emulsion bubbles \cite{Tchuenbou-Magaia2009}.

In a recently published experiment \cite{Knoche2013}, the deformation of a
hydrophobin-coated bubble rising from a capillary was investigated. As
a result of deflating the bubble, wrinkles appeared on the bubble surface
proving that the protein layer has elastic properties and a nonvanishing
shear modulus. Furthermore, 
the analysis of this experiment  showed that the
elastic response is very nonlinear, and a steep increase in the measured
elastic modulus with increasing bubble deflation was observed. It was
conjectured that this steep increase is related to the molecular structure of
the hydrophobin proteins consisting of a hard core (a $\beta$ barrel) with a
softer shell (loop and coil structures). The hard cores coming into contact
could then be the cause for the steep increase in the elastic modulus.

\begin{figure}
 \centering
 \includegraphics{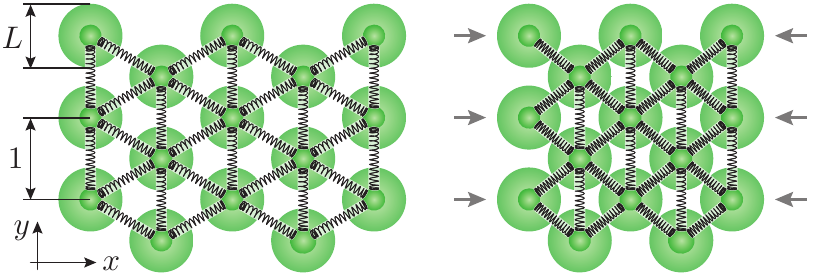}
 \caption{ Bead--spring model with hard cores for interacting hard particles at
   the air/water interface or a liquid interface.  The springs give the
   network a Hookean elasticity, and the hard cores (green disks, diameter
   $L$) impose limits on the maximal admissible compression (see configuration
   on the right). The lattice constant is normalized to $1$ and $L<1$.  }
\label{fig:Federnetzwerk}
\end{figure}

In order to verify this interpretation, we develop an elasticity model based
on the microscopic view of a particle raft 
sketched in Figure~\ref{fig:Federnetzwerk}: Globular
particles interact by soft springs (corresponding to an outer soft shell) and
steric interactions (corresponding to hard cores). From this microscopic view,
we derive continuum elastic laws, which can then be used to 
 analyze the deflated shapes of capsules with this type of elasticity.
The hypothesis underlying this approach is the following: The steep increase
in the surface Young or stretch modulus that was obtained in
ref~\citenum{Knoche2013} is a result of applying linear Hookean elasticity
in a situation where the
proposed elasticity model for the protein particle raft that includes 
hard cores should be more appropriate. Therefore, 
 we should be able to describe the observed  
shapes of hydrophobin-coated bubbles more exactly and 
with much less variation of the elastic moduli along the deflation 
trajectory if the new hard-core/soft-shell
elasticity model is used. 

The proposed elasticity model is quite generic: It is a generalization of a
standard bead--spring model for 
Hookean elastic membranes \cite{Seung1988}, which also takes into 
account  hard cores of the beads. 
On the other hand, it generalizes a particle raft elasticity model
presented in ref~\citenum{Vella2004}, 
which considers exclusively hard-core interactions.
Therefore, our hard-core/soft-shell
elasticity model applies  more widely than to
monolayers of the protein hydrophobin.
We expect such contact interaction to be relevant for many 
liquid interfaces decorated with interacting hard particles \cite{Zeng2006},
ranging 
from colloidal rafts covering colloidosomes \cite{Dinsmore2002} or 
``armoured droplets'' \cite{Subramaniam2005} to rafts of 
larger particles \cite{Vella2004}.

Our results will show that 
the elastic properties of the two-dimensional material 
are Hookean for small compression, where only soft springs are loaded 
without hard-core contact but strongly modify 
as soon as compression brings the hard cores of particles 
into contact. 
The constitutive stress--strain 
relations of the soft spring, which 
 determine the material properties for small compression, have to be 
properly connected to stress relations, which are governed by 
force balance for compressions that induce hard-core contact. 
The transition between both regimes can take place between different 
regions within  the same material, as we will see in the analysis 
of shapes of pendant or buoyant capsules attached to a capillary. 
A detailed and quantitative 
 understanding of shapes and deformations of such compressed 
particle-decorated liquid interfaces therefore requires a model 
as presented here, which 
properly connects Hookean elasticity of soft springs or soft particle
interactions with  hard-core elasticity,  and goes beyond 
a simple  effective description based on  Hookean elasticity.

\section{Continuum Description of a Bead--Spring Model with Hard Cores}

\begin{figure*}
 \centering
 \includegraphics{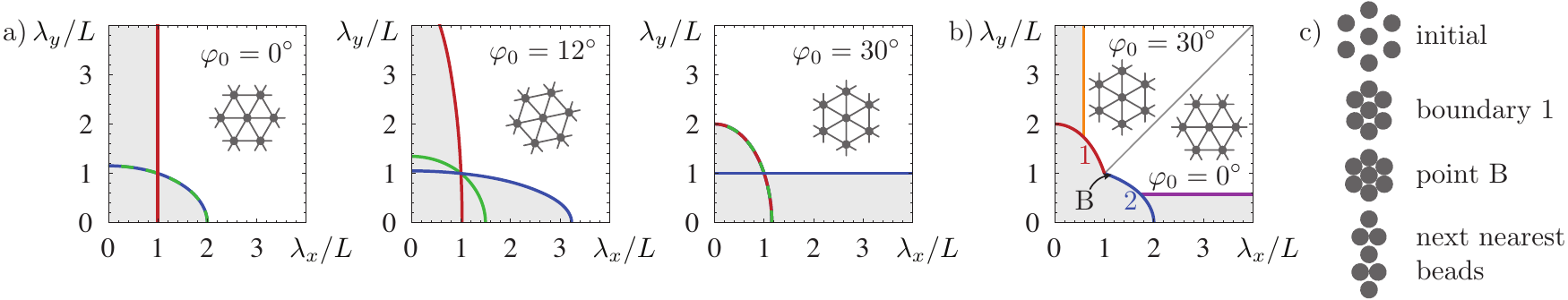}
 \caption
 { Forbidden domains (light gray) in the $(\lambda_x, \lambda_y)$-plane
   limited by ellipses.  (a) Different orientations $\phi_0$ produce different
   forbidden domains. Straight lines are degenerate ellipses with infinite
   major axis, and dashed lines with alternating color indicate if two of the
   three ellipses are identical. The pictograms show the lattice orientation.
   (b) Choice of $\phi_0$ that minimizes the area of the forbidden domain. For
   easy identification, the boundaries are labeled 1 and 2; and the point on
   both lines is termed B.  There is an additional vertical and horizontal
   boundary at $\lambda_i/L = 1/\sqrt{3}$ where next nearest beads come into
   contact.  (c) Sketches of the different possibilities of contact between the
   beads, here for $\phi_0=30^\circ$, that is,  when the compression is
   predominantly in $x$-direction.
}
 \label{fig:ellipses}
\end{figure*}

We assume that the beads are arranged in the $(x, y)$-plane in a hexagonal
crystal with a lattice constant normalized to $1$, see
Figure~\ref{fig:Federnetzwerk}. Such an arrangement is the closest packing of
spheres and it behaves isotropically \cite{Landau1987}, so that the continuum
elasticity of the membrane can be characterized by two elastic moduli, for
example the surface Poisson ratio $\nu$ and surface Young modulus 
$\EHo$. On the
microscopic level, the elastic response of the membrane is governed by the
spring constant $k$. By evaluation of the deformation energy within one unit
cell, it can be shown that the continuum elastic moduli are determined by
\cite{Ostoja-Starzewski2002, Seung1988}
\begin{equation}
 \EHo = \frac{2}{\sqrt{3}} \, k \qq{and} \nu = \frac{1}{3}.
\end{equation}
Without the hard-core interactions, the membrane can thus be described with
the usual Hookean elasticity as specified by the strain-energy density
\cite{Libai1998}
\begin{equation} \label{eq:w_S_pend}
\begin{multlined}
 w_S(\lambda_x, \lambda_y) = 
\frac{1}{2} \frac{\EHo}{1-\nu^2} \big[ (\lambda_x-1)^2 \\
   + 2 \nu (\lambda_x-1)(\lambda_y-1) 
 + (\lambda_y-1)^2 \big] + \lambda_x \lambda_y \gamma,
\end{multlined}
\end{equation} 
where $\lambda_x$ and $\lambda_y$ denote the stretches in $x$- and
$y$-direction, respectively.
The strain energy density is measured per surface area of the undeformed
interface. The form of the surface energy density in eq~\ref{eq:w_S_pend} 
is valid only for
deformations where the $x$- and $y$-directions are the principal directions of
the strain tensor, which we assume throughout this paper. This means that a
line element $(\diff x_0, \diff y_0)$ is mapped to $(\diff x, \diff y) =
(\lambda_x \diff x_0, \lambda_y \diff y_0)$ by the deformation.

The term $\lambda_x \lambda_y \gamma$ accounts for an isotropic surface
tension $\gamma$ acting in the liquid surface \cite{Knoche2013}.  We assume
that the corresponding liquid interface covers the entire deformed area
$\lambda_x \lambda_y A_0$, where $A_0$ is the undeformed reference interface
area.  
In the presence of adsorbed particles, which are only partially wet, a
fixed area $A_{\rm hard}$ of the liquid surface is replaced by a
liquid-solid-interface with a surface tension $\gamma_{\rm hard}$, 
and the last term in eq~\ref{eq:w_S_pend} becomes
$(\lambda_x \lambda_y - A_{\rm hard}/A_0) \gamma + (A_{\rm hard}/A_0)
\gamma_\text{hard}$.
This shift by a constant independent of the  stretching factors does 
neither change the interface stresses nor the resulting shape equations.

From the energy density eq~\ref{eq:w_S_pend}, 
the stresses acting in the interface can be derived \cite{Libai1998}.
 A superscript $^{(\text{s})}$ indicates that this is the
contribution of the springs,
\begin{equation} \label{eq:taus}
 \taus_x(\lambda_x, \lambda_y) = \frac{\EHo}{1-\nu^2} \frac{1}{\lambda_y} 
   \left[ (\lambda_x - 1) + \nu (\lambda_y - 1) \right] + \gamma,
\end{equation} 
and with indices $x$ and $y$ interchanged for $\taus_y$.
We note that compression with $\lambda_i<1$ gives rise to negative 
elastic contributions to the stresses, whereas 
the surface tension $\gamma$ always provides  a positive 
contribution, such that $\tau_i\le \gamma$. 

Now we have to evaluate how these results from the spring elasticity 
are modified by  the steric interactions between the
hard cores. The springs in the lattice have a rest length of $1$ and can be
oriented along three different directions $i$, which we characterize by the
angle $\phi_i$ to the $x$-axis. This angle can take the values $\phi_i =
\phi_0 + i \pi/3$, with $i \in \{-1, 0, 1\}$, where $\phi_0$ determines the
overall orientation of the lattice in the $(x, y)$-plane.
Upon deformation, the length of a spring along direction $i$ changes from $1$
to
\begin{equation}
 d_i = \left| \zweiervec{\lambda_x \cos \phi_i}{\lambda_y \sin \phi_i} \right| 
    = \sqrt{ \lambda_x^2 \cos^2 \phi_i + \lambda_y^2 \sin^2 \phi_i }.
\end{equation}
The steric interactions enforce that the springs of the lattice can be
compressed at maximum to a minimal length of $L$ (with $L<1$),
which corresponds to the diameter of the hard cores,
measured in units of the lattice constant.
Thus, we have three conditions $d_i \geq L$ to be satisfied, or
equivalently
\begin{equation}
 \begin{aligned}
   \lambda_x^2 \, \cos^2( \phi_0 - \pi/3) + \lambda_y^2 \, 
        \sin^2( \phi_0 - \pi/3) &\geq L^2 \\
   \lambda_x^2 \, \cos^2( \phi_0 ) + \lambda_y^2 \, 
        \sin^2( \phi_0 ) &\geq L^2 \\
   \lambda_x^2 \, \cos^2( \phi_0 + \pi/3) + \lambda_y^2 \, 
       \sin^2( \phi_0 + \pi/3) &\geq L^2.
 \end{aligned}
\end{equation} 
In the $(\lambda_x, \lambda_y)$-plane, these conditions specify three ellipses
to be excluded from the admissible domain for the stretches; see
Figure~\ref{fig:ellipses}a. All three ellipses contain the point $(L, L)$.

The excluded or ``forbidden'' domain of stretches, 
shaded light gray in Figure~\ref{fig:ellipses}a and b, depends on the
orientation $\phi_0$ of the lattice. We choose this parameter according to the
following rule: If $\lambda_x < \lambda_y$, then $\phi_0 = 30^\circ$;
otherwise $\phi_0 = 0^\circ$.  With this choice, the forbidden domain becomes
as small as possible; see Figure~\ref{fig:ellipses}b. Choosing $\phi_0$ in
dependence of the stress state means that it may change during a deformation;
 for particle rafts this seems plausible because the particles are not
rigidly cross-linked, but may rearrange  to change their lattice orientation
 from a $0^\circ$ 
 to a $30^\circ$ state (see pictograms in Figure~\ref{fig:ellipses}c). 
If the stress state of the membrane becomes inhomogeneous, 
as it will happen, for example, for capsules attached to a 
capillary as considered below in more detail, we may 
encounter regions of $0^\circ$ and $30^\circ$
orientation within a single membrane. Then,  
problems might arise at the boundary
separating two regions of different orientation, because the lattices cannot
be joined properly. This could give rise to a line energy; we neglect
these complications in the following.

With our choice of $\phi_0$, the boundary line  between the forbidden and
admissible domain of the stretch plane can be described by 
 $(\lambdab_x, \lambdab_y ) = (\lambdab_x (\lambda_y), \lambda_y )$
with 
\begin{equation} \label{eq:lambda_boundary}
  \lambdab_x (\lambda_y) = 
    \begin{cases}
      \sqrt{4 L^2 / 3 - \lambda_y^2 / 3} & 
     \text{if } \lambda_y > L \text{ (boundary 1)} \\
      \sqrt{4 L^2 - 3 \lambda_y^2} & 
      \text{if } \lambda_y < L \text{ (boundary 2)}.
    \end{cases}
\end{equation}
Here we have introduced the terms boundary 1 and 2, which have to be
distinguished in most of the following calculations. They are plotted in red
and blue, respectively, in Figure~\ref{fig:ellipses}b. The point $(\lambda_x,
\lambda_y) = (L, L)$ which is on both boundary lines is termed point B.

If the external loads try to push the membrane into the forbidden domain, the
hard-core interactions keep it on the boundary by
providing additional contributions to the stresses $\tau_x$ and
$\tau_y$. These hard-core contributions are denoted by $\tauc_x$ and
$\tauc_y$. The complete stresses then read
\begin{equation} \label{eq:stresses_complete}
 \tau_x = \taus_x(\lambda_x, \lambda_y) + \tauc_x \qq{and}
 \tau_y = \taus_y(\lambda_x, \lambda_y) + \tauc_y
\end{equation} 
with $(\lambda_x, \lambda_y)$ being a point on the boundary of the forbidden
domain as given by eq~\ref{eq:lambda_boundary}. 
Since $\tauc_x$ and $\tauc_y$ are
transmitted through the ``skeleton'' of hard cores, they must satisfy certain
conditions 
of force equilibrium 
that can be derived from the geometry of the lattice. In Appendix 
\ref{app:tauc_ratios},
it is shown that force balance 
prescribes the ratio of the hard-core stresses to
\begin{equation} \label{eq:tauc_ratios}
\begin{aligned}
 \frac{\tauc_y}{\tauc_x} = 
  \begin{cases}
   {\lambda_y^2}/3{\lambda_x^2} & 
        \text{if } \lambda_y > L \text{ (boundary 1)} \\
   3 {\lambda_y^2}/{\lambda_x^2} & 
       \text{if } \lambda_y < L \text{ (boundary 2)}
  \end{cases} \\
 \text{and} \quad \frac{1}{3} \leq \frac{\tauc_y}{\tauc_x} \leq 3 \quad 
       \text{if } \lambda_x = \lambda_y = L \text{ (point B)}.
\end{aligned}
\end{equation} 
Note that $(\lambda_x, \lambda_y)$ must always be a point on the boundary of
the forbidden domain when these equations are applied to calculate the
hard-core contributions to the stresses.  

For strong compressive strains with 
$\lambda_x<1/\sqrt{3}$ or $\lambda_y<1/\sqrt{3}$,
next nearest beads, which are not connected by springs, 
come into contact; see Figure~\ref{fig:ellipses}b, orange and violet lines.
Such compressive strains are not relevant for the applications discussed
in this Article.

\begin{figure}
 \centering
 \includegraphics{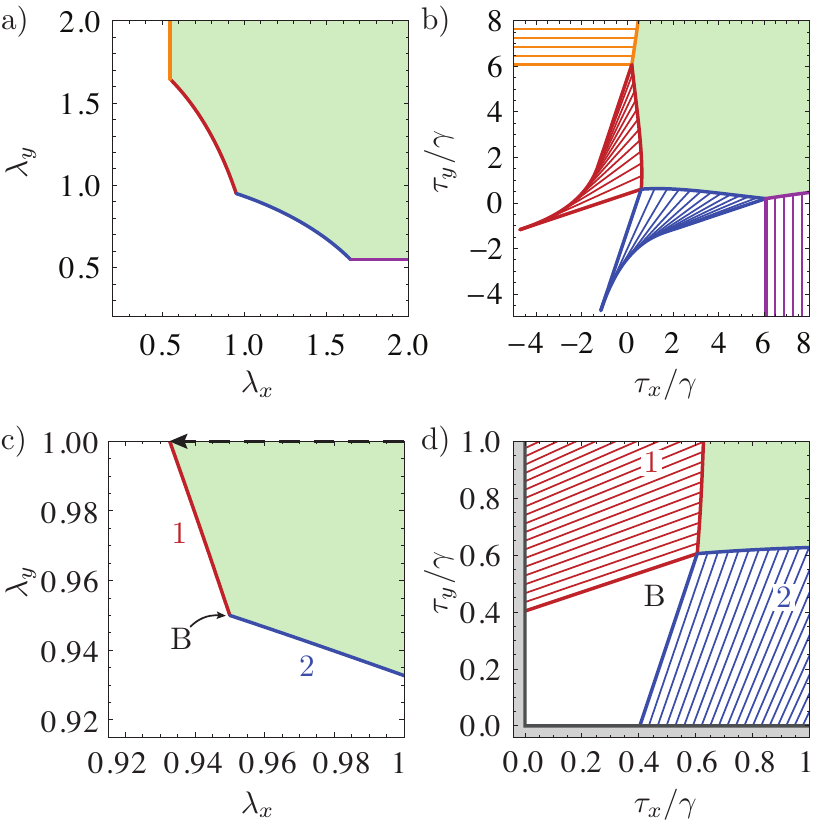}
 \caption{
     Admissible domains (light green) and hard-core interaction
     domains for stretches (a, c)   and stresses (b, d) for $L=0.95$
     and $\EHo = 5 \gamma$. The top row shows large views of the stress and
     stretch planes, and the bottom row close-ups of the relevant region. In
     (b) and (d), the straight thin lines indicate the stresses that can be
     reached from a point on the boundary by adding the hard-core
     contributions to the elastic stresses. 
       The wrinkling region is shaded in
     gray in (d). 
  The arrow in (c) indicates the 
  trajectory of uniaxial compression of a planar layer considered 
    below.
}
\label{fig:schmetterlinge}
\end{figure}

The above results for our custom elasticity model for particle rafts are
summarized and illustrated in Figure~\ref{fig:schmetterlinge}.  
Figure~\ref{fig:schmetterlinge}a and b
shows a large view of the stretch plane and stress plane with different
regions. Figure~\ref{fig:schmetterlinge}c and d focuses
  on the regime relevant for compressed
membranes: $\lambda_i \leq 1$ and $0 < \tau_i \leq \gamma$. The light green
regions are the admissible regions, where the hard cores are not in
contact. Here, the usual Hookean elasticity (eq~\ref{eq:taus}) is valid, and
there is a bijective mapping between $(\lambda_x, \lambda_y)$ points in the
stretch plane and $(\tau_x, \tau_y)$ points in the stress plane.

On boundaries 1 and 2, the hard cores come into contact. In the stretch plane,
this boundary cannot be trespassed because of the geometric constraints 
imposed by the undeformable hard cores: Even if the external forces try to push
the lattice beyond this line, the lattice will get stuck on the boundary of
the forbidden domain. In the stress plane, however, the points beyond
boundaries 1 and 2 can be accessed by including the hard-core contributions
$\tauc_x$ and $\tauc_y$ in the stresses because hard cores can transmit
force even though they are undeformable.  

A point $\big(\lambdab_x, \lambdab_y \big)$ on the boundary in the stretch
space is mapped by Hooke's law to a point $\big( \taub_x, \taub_y \big)$ on
the boundary in the stress space. From this point on, stresses $(\tau_x,
\tau_y) = \big( \taub_x, \taub_y \big) + \big( \tauc_x, \tauc_y \big)$ can be
reached, where the hard-core contributions $\tauc_x$ and $\tauc_y$ must be
negative (because the skeleton can only support compressive stresses) and must
obey the force balance constraint (eq \ref{eq:tauc_ratios}). The submanifold of
stresses which is accessible from the point on the boundary is, thus, a
straight line with a slope $\lambda_y^2/3\lambda_x^2$ starting
from boundary 1 or $3\lambda_y^2/\lambda_x^2$ starting from
boundary 2 in Figure~\ref{fig:schmetterlinge}d  (see the thin red and blue
lines).  
The analogous discussion of the boundaries where next nearest beads
come into contact is spared because it is not relevant for the applications
presented below.

For point B, with $\lambda_x = \lambda_y = L$, the ratio of the hard-core
stresses is not fixed to a certain value. Instead, it can range from $1/3$ to
$3$. In the plot of the stress plane, Figure~\ref{fig:schmetterlinge}d, this
means that the whole white region is accessible from point B.

\section{Compression of Planar Films}
\label{sec:planar}

To get a first impression of the elastic behaviour of a 
particle layer with hard cores
at a fluid interface, we investigate a typical compression experiment in a
Langmuir trough \cite{Cicuta2009,Aumaitre2011}.
In this setup, a monolayer is spread on a
water surface and subsequently it is compressed in $x$-direction by moving the
barriers of the trough. A Wilhelmy plate can then measure the stresses
$\tau_x$ and $\tau_y$.

If we consider $\lambda_y = 1$ as fixed and compress the layer with a ratio
$\lambda_x < 1$, we expect the monolayer to wrinkle at sufficiently high
compression. The determination of the critical compression is a purely
geometric problem. In the stretch space of Figure~\ref{fig:schmetterlinge}c, 
we follow a horizontal line until we cross boundary 1 
as indicated by the  arrow. Then the layer must
wrinkle because the hard cores cannot be compressed.

On the other hand, if we consider the compressive stress $\tau_x < 0$ to be
given, we can determine the critical stress from the stability equation
\cite{Vella2004,Landau1970}
\begin{equation} \label{eq:stab}
 E_B \partial_x^4 w(x) - \tau_x \partial_x^2 w(x) + \rho g w(x) = 0,
\end{equation} 
where $E_B$ is the bending stiffness of the layer, $w(x)$ is the displacement in
the $z$-direction which is assumed to be independent of $y$, and $\rho g$ is the
fluid density times acceleration of gravity. 
The bending rigidity $E_B$ can arise from 
 an additional  bending stiffness of connecting springs.
For hydrophobin layers this corresponds to a bending rigidity of the 
soft shell of the protein. 
Assuming sinusoidal wrinkles of
the form $w \sim \sin kx$, eq~\ref{eq:stab} gives a critical stress
$\tau_x$ which still depends on the wavenumber $k$ of our ansatz. Minimizing
the magnitude of the stress with respect to $k$ gives a critical wavelength
and critical stress of \cite{Vella2004}
\begin{equation}
 \lambda_c = 2\pi \left( E_B/\rho g \right)^{1/4} \quad \text{and} 
   \quad \tau_{x,c} = - 2 \sqrt{E_B \rho g}.
\label{eq:lambdac}
\end{equation} 
This is the same result as that for an elastic membrane without hard cores.
The
 result for $\lambda_c$ differs from what has been obtained in
  ref~\citenum{Vella2004}: we do not find a dependence on the packing fraction
  of hard particles because the bending modulus corresponding to the
  soft springs appears in eq~\ref{eq:lambdac} rather than an effective
  modulus depending on packing fraction as in ref~\citenum{Vella2004}.

\begin{figure}
 \centering
 \includegraphics{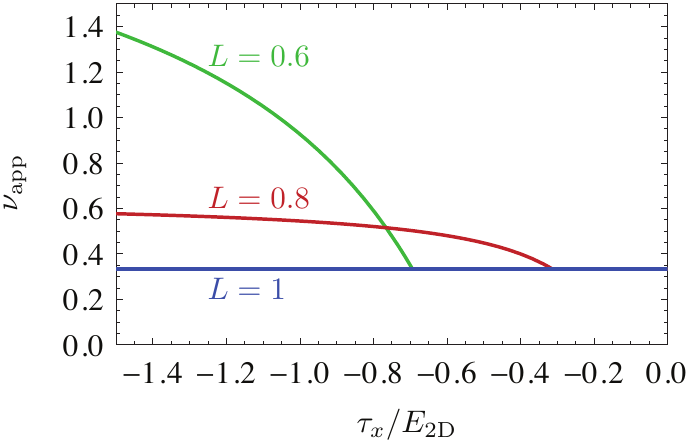}
 \caption{
 Apparent Poisson ratio $\nu_\text{app} = {\Pi_y}/{\Pi_x}$ in our
   elasticity model for different values of $L$ as indicated. At the kinks,
   the hard cores come into contact and the hard-core contributions to the
   stresses begin to influence this ``measured'' Poisson ratio.}
 \label{fig:nu_app}
\end{figure}

Differences between membranes with and without hard cores can be found when
the Poisson ratio of the layer is measured. For uniaxial compressive 
strain in the $x$-direction ($\lambda_x<1$, $\lambda_y=1$), 
the Poisson ratio of a linear elastic material can be measured
as $\nu = \tau_y/\tau_x$. 
Now let us apply this ``measurement'' to the more
complex total stresses in eq~\ref{eq:stresses_complete}:
\begin{equation}
\nu_\text{app}  = \frac{\taus_y+\tauc_y-\gamma}{\taus_x+\tauc_x-\gamma}
   = \frac{\Pi_y}{\Pi_x}.
\end{equation}
Here, we 
subtract the fluid surface tension $\gamma$ from the 
total stresses, which provides a  background tension and is typically 
subtracted in  measurements of the surface pressures 
$\Pi_x$ and $\Pi_y$ \cite{Cicuta2009,Aumaitre2011}.
For simplicity,
we also neglect the $1/\lambda_x$ prefactor in relation 
(\ref{eq:taus}) for $\taus_y$, which
represents a geometrical nonlinearity. In the regime where the hard cores are
in contact, the apparent Poisson ratio thus reads
\begin{equation} \label{eq:nuapp}
 \nu_\text{app} = \frac{\frac{\nu}{1-\nu^2} (\lambdab_x - 1) 
+ \ttauc_x /3 (\lambdab_x)^2 }{\frac{1}{1-\nu^2} (\lambdab_x - 1) + \ttauc_x}
\end{equation} 
with $\lambdab_x = \sqrt{4L^2/3-1/3}$ and $\ttauc_x = \tauc_x/\EHo$. The hard
cores have a nonvanishing contribution $\ttauc_x$ only if the applied
compressive stress satisfies $\tau_x < \frac{\EHo}{1-\nu^2} (\lambdab_x - 1)$
so that the hard cores are in contact. For larger $\tau_x$ or weak
compression,
 the usual Hookean law is valid,
 which leads to $\nu_\text{app} = \nu = 1/3$. In the limit 
of strong compression where
the hard-core contribution dominates the stresses, we have $\nu_\text{app} =
1/3(\lambdab_x)^2 = 1/(4L^2 - 1)$, which is always larger than $1/3$ for $L<1$
and can even become larger than $1$. 
For $L<1/\sqrt{3}$, this relation changes because next 
nearest beads come into contact. 
This shows that contact of hard cores can give  rise to a more effective 
redirection of compressive stress into the perpendicular direction,
which gives rise to a stress-dependent  Poisson ratio increasing with
compression. 
The crossover between  weak compression with  the
 Hookean  value $\nu = 1/3$ and the 
increasing apparent Poisson ratio $\nu_\text{app}$ 
for strong compression is shown in Figure~\ref{fig:nu_app} for different
values of $L$. 
Our results differ from $\nu_\text{app} = 1/\sqrt{3}$, which 
has been found in ref~\citenum{Vella2004} 
for pure hard-core particle rafts.
In ref~\citenum{Clegg2008}, on the other hand, 
it has been argued that $\nu_\text{app} = 1/3$
is correct also for pure hard-core particle rafts. 
Our  result (eq \ref{eq:nuapp}) for the more general elasticity 
model explains that, 
for rafts of interacting hard particles, where the interaction provides 
a ``soft shell'', experimental measurements such as those in 
refs~\citenum{Cicuta2009} and \citenum{Aumaitre2012}  
should be interpreted using a 
stress-dependent Poisson ratio, which  increases beyond 
 $1/3$ and up to $\nu_\text{app}  = 1/(4L^2 - 1)$  
if hard cores come into contact; see Figure~\ref{fig:nu_app}.

In ref~\citenum{Aumaitre2012}, for example, 
an experiment is reported where a monolayer of
hydrophobin molecules is compressed in a Langmuir trough. Two Wilhelmy plates
were used to measure the stresses in $x$- and $y$-directions, respectively, and
from the data one can calculate an apparent Poisson ratio of $\nu_\text{app}
\approx 0.6-0.7$. This relatively large value can be well explained with
our elasticity model 
if $L \lesssim 0.8$ according to Figure~\ref{fig:nu_app}.
It is tempting to compare this hard-core length to available 
molecular information. In ref~\citenum{Aumaitre2011}, it has been 
observed that visual buckling, which could be due to hard-core 
contact, happens at a molecular area of $a_{\rm buckle} \approx 
347 \, {\mbox{\normalfont\AA}}^2$ 
per HFBII hydrophobin 
protein. Assuming closed-packed hard cores, this corresponds to 
a diameter of $20 \, {\mbox{\normalfont\AA}}$, 
if HFBII is assumed to be spherical.
A value $L=0.8$ then corresponds to a total  diameter of 
$25 \, {\mbox{\normalfont\AA}}$
of the protein. This value is consistent, for example, 
with the dimensions 
$24 \times 27 \times 30 \, {\mbox{\normalfont\AA}}$ given for 
 the whole HFBII molecule as obtained from X-ray
 crystallography \cite{Hakanpaa2004}.
 Grazing-incidence X-ray diffraction and reflectivity results 
in ref~\citenum{Kisko2009} hint, however, 
 at smaller diameters 
$20-24 \, {\mbox{\normalfont\AA}}$ and, thus, 
larger values of $L$.
In refs~\citenum{Aumaitre2012} and \citenum{Aumaitre2012b}, 
plateaus in the surface pressure isotherms 
have been observed, which can be interpreted as a liquid--gas coexistence of 
a gas phase of isolated hydrophobins and a dilute liquid where soft shells
of hydrophobins just start to make contact. 
Taking the area per molecule $a_{\rm plateau}$ 
at the high-density end of these plateaus 
we find  values of $L=(a_{\rm plateau}/a_{\rm buckle})^{1/2}$ in the range 
$L=0.85-0.90$.
Overall, a reliable estimate of $L$ based on  molecular information 
appears difficult.

\section{Shape Equations in the Pendant/Rising Capsule Geometry}

\begin{figure}
 \centering
 \includegraphics{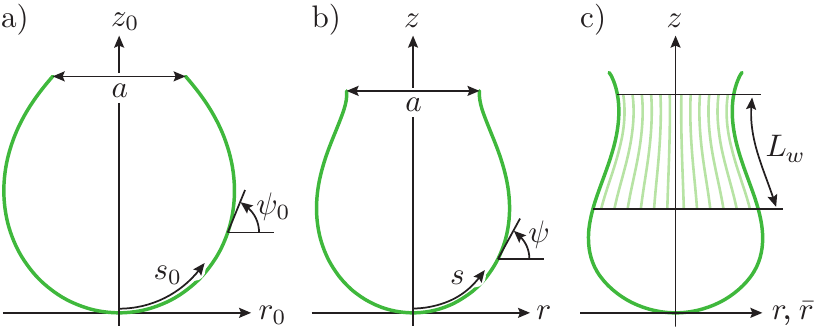}
 \caption{Arc-length parametrizations in cylindrical coordinates of the
   undeformed $(r_0(s_0), z_0(s_0))$, deformed $(r(s), z(s))$ and the wrinkled
   midsurface. The wrinkled region of length $L_w$ is described by an
   axisymmetric pseudosurface around which the real midsurface oscillates.}
 \label{fig:parametrisations}
\end{figure}

We now apply the previously developed elasticity model to the shape equations
for axisymmetric 
capsules attached to a capillary which have been developed in
ref~\citenum{Knoche2013}. 
This is a challenging problem because it turns out that
the transition between the  elasticity regimes A, 1, 2, and B 
can take place within  the same material
between different axisymmetric regions 
 along the capsule. Therefore, we have to generalize the shape 
equations from ref~\citenum{Knoche2013} to  include switching 
between different constitutive elastic laws corresponding to 
the Hookean regime A and the hard-core dominated regimes 1, 2, and B 
along the capsule contour. 

The shape equations are derived from non-linear shell theory and their
solution describes the shape and stress distribution of a deformed pendant or
rising capsule. Figure~\ref{fig:parametrisations} shows the parametrization of
the undeformed and deformed capsule shape. At its upper rim, the capsule is
attached to a capillary of diameter $a$. The reference shape (see
Figure~\ref{fig:parametrisations}a) is a solution of the Laplace-Young
equation with an isotropic interfacial tension $\gamma$ and a density
difference $\Delta \rho$ of inner and outer fluid \cite{Knoche2013}. This
shape can be deformed, for example by reducing the internal capsule
volume. Each point $(r_0(s_0), z_0(s_0))$ is mapped onto a point $(r(s_0),
z(s_0))$ in the deformed configuration, which induces a meridional and
circumferential stretch, $\lambda_s = \diff s/\diff s_0$ and $\lambda_\phi =
r/r_0$, respectively. The arc length element $\diff s$ of the deformed
configuration is defined by $\diff s^2 = (r'(s_0)^2 + z'(s_0)^2) \diff s_0^2$.

The shape equations that determine the deformed configuration 
describe  force and torque balance  within the capsule shell and 
are given by \cite{Knoche2013}
\begin{equation} \label{eq:shape_pend}
 \begin{aligned}
  r'(s_0) &= \lambda_s \cos \psi\\
   z'(s_0) &= \lambda_s \sin \psi\\
  \psi'(s_0) &= \frac{\lambda_s}{\tau_s} 
   (p - \Delta \rho \, g \, z - \kappa_\phi \tau_\phi) \\
  \tau_s'(s_0) &= \lambda_s \frac{\tau_\phi - \tau_s}{r} \cos \psi.
 \end{aligned}
\end{equation} 
Here, $\psi$ denotes the slope angle as defined in
Figure~\ref{fig:parametrisations}, $\tau_s$ and $\tau_\phi$ are the meridional
and circumferential stress, respectively, $\kappa_\phi = \sin \psi/r$ is the
circumferential curvature, and $p-\Delta\rho \, g \, z$ is the hydrostatic
pressure difference exerted on the capsule membrane. In this formulation, the
bending stiffness has been neglected since the typical capsules used for these
experiments are very thin and bendable \cite{Knoche2013}.

In order to solve this system of ordinary differential equations numerically
with a shooting method, the quantities $\lambda_s$ and $\tau_\phi$ occurring
on the right-hand side of the system must be calculated for given
$\lambda_\phi$ and $\tau_s$ by virtue of an elastic constitutive law. For the
simple Hookean elastic law in eq \ref{eq:taus} this was demonstrated in
ref~\citenum{Knoche2013},
\begin{equation} \label{eq:closing}
 \begin{aligned}
  \lambda_s &= \frac{1-\nu^2}{ \EHo } \, \lambda_\phi \, 
        (\tau_s - \gamma) - \nu (\lambda_\phi -1) +1, \\
  \tau_\phi &= \frac{ \EHo }{1-\nu^2} \, \frac{1}{\lambda_s} 
      \big( (\lambda_\phi-1) + \nu\, (\lambda_s -1) \big)+ \gamma.
 \end{aligned}
\end{equation} 
The boundary conditions for the system of shape equations are $r(0) = z(0) =
\psi(0) = 0$ and $r(L_0) = a/2$, where $L_0$ is the contour length in the
undeformed configuration and the end-point of the integration. The starting
value $\tau_s(0)$ is free and serves as a shooting parameter to satisfy the
boundary condition at $L_0$ \cite{Knoche2013}.

 Here, we want to use the elasticity model  for
 particle rafts developed above (eqs \ref{eq:taus}, \ref{eq:stresses_complete}
 and \ref{eq:tauc_ratios}), including the hard-core interactions. 
 Therefore, depending on the size of compressive stresses, 
 we have to switch between different  constitutive elastic
 relations corresponding to Hookean  constitutive relations 
 or hard-core  constitutive relations  
 along the arc-length coordinate $s_0$ as explained in detail in 
Appendix \ref{app:II}.

\subsection{Wrinkling and Modified Shape Equations}

In our model, we neglect the bending stiffness of the capsule membrane. The
fact that the membrane is infinitely easy to bend implies that it will
immediately wrinkle under compressive stresses. 
In ref~\citenum{Knoche2013}, it
was shown that deflated pendant capsules exhibit wrinkles due to a compressive
hoop stress $\tau_\phi < 0$, as also indicated in
Figure~\ref{fig:parametrisations}c. 
The capsule membrane is not axisymmetric
in the wrinkled region, but can be approximated by an axisymmetric
pseudosurface around which the real membrane oscillates. 
The shape of this pseudosurface is determined by setting $\tau_\phi =
0$ where the original shape equations predict negative hoop stresses. 
The assumption $\tau_\phi = 0$ on the
pseudosurface is common to various theories of wrinkling, for example
 tension field theory \cite{Steigmann1990} or  far-from-threshold
theory \cite{Davidovitch2011}.
This
leads to a modified system of shape equations
\begin{equation} \label{eq:shape_wrink}
 \begin{aligned}
  \bar r'(s_0) &= \lambda_s \cos \psi  \\ 
   z'(s_0) &= \lambda_s \sin \psi \\
  \psi'(s_0) &= \frac{\lambda_s}{\bar \tau_s} (p - \Delta \rho \, g \, z) \\
  \bar \tau_s'(s_0) &= - \lambda_s \frac{\bar \tau_s}{\bar r} \cos \psi
 \end{aligned}
\end{equation} 
where $\bar r$ is the radial coordinate of the pseudo-surface and $\bar
\tau_s$ the meridional stress measured per unit length of the pseudosurface,
that is $\bar \tau_s = \tau_s \lambda_\phi/\bar\lambda_\phi$ when $\bar
\lambda_\phi = \bar r/r_0$ denotes the pseudo hoop stretch. The system is
closed by the equation
\begin{equation} \label{eq:closing_wrink}
 \lambda_s = \frac{\bar \tau_s \bar \lambda_\phi + \EHo 
    - \gamma(1+\nu)}{\EHo - 2\nu \gamma - \gamma^2 (1-\nu^2)/\EHo}.
\end{equation} 
A discussion of these modified shape equations for the pseudo-surface and the
automatic switching between normal and modified shape equations during the
integration can be found in ref~\citenum{Knoche2013}.

Wrinkling can also occur when the lattice is jammed on
 boundary 2 or point B. As
in ref~\citenum{Knoche2013}, 
we handle wrinkling by introducing a pseudosurface
(indicated with an overbar) and setting the hoop stress to zero. On boundary
2, we have $\lambda_\phi = \sqrt{4L^2/3 - \lambda_s^2/3}$ (note that
$\lambda_\phi$ refers to the hoop stretch of the real, wrinkled surface and is
to be distinguished from the pseudo hoop stretch $\bar \lambda_\phi = \bar r/
r_0$). The condition $\tau_\phi = 0$ is equivalent to
\begin{equation}
 \tauc_\phi = - \gamma - \frac{\EHo}{1-\nu^2} \frac{1}{\lambda_s} 
  \left[ \big(\lambda_\phi -1 \big) + \nu (\lambda_s - 1) \right].
\end{equation} 
We further have $\tauc_s = \tauc_\phi \cdot \lambda_s^2 / 3
\lambda_\phi^2$. So the complete meridional tension, measured per unit length
of the pseudosurface, is determined by
\begin{equation}
 \bar\tau_s = \frac{\lambda_\phi}{\bar \lambda_\phi} 
  \left[ \taus_s(\lambda_s, \lambda_\phi) + 
     \frac{1}{3} \frac{\lambda_s^2}{\lambda_\phi^2} \, \tauc_\phi \right].
\end{equation} 
This is a quite complicated function of $\lambda_s$, because $\tauc_\phi$ and
$\lambda_\phi$ herein also depend on $\lambda_s$. It must be solved for
$\lambda_s$ to evaluate the right-hand side of the shape equations
(eq \ref{eq:shape_wrink}), 
which is done numerically in each integration step of
the shape equations.

When wrinkling occurs in point B, the shape equations for the pseudosurface
(eq \ref{eq:shape_wrink}) can also be used, and since $\lambda_s = \lambda_\phi =
L$, the inversion of a stress--strain relation to obtain $\lambda_s$ is not
necessary. In point B, the hard-core stresses $\tauc_s$ and $\tauc_\phi$ are
independent of each other, and the spring contributions $\taus_s$ and
$\taus_\phi$ are fixed because of $\lambda_s = \lambda_\phi = L$. From the
wrinkling condition $\tau_\phi = 0$ we can, thus, calculate $\tauc_\phi$. The
meridional hard-core contribution $\tauc_s$ must be calculated from the
differential equation for $\tau_s$. Note that the geometry of the
pseudosurface is not fixed by the condition $\lambda_s = \lambda_\phi = L$
because the circumferential stretch $\bar \lambda_\phi$ of the pseudosurface
is free. This is in contrast to the case of a lattice being stuck in point B
without wrinkling as discussed above.

\subsection{Numerical Integration and Switching between the Shape Equations}

\begin{figure}
 \centering
 \includegraphics{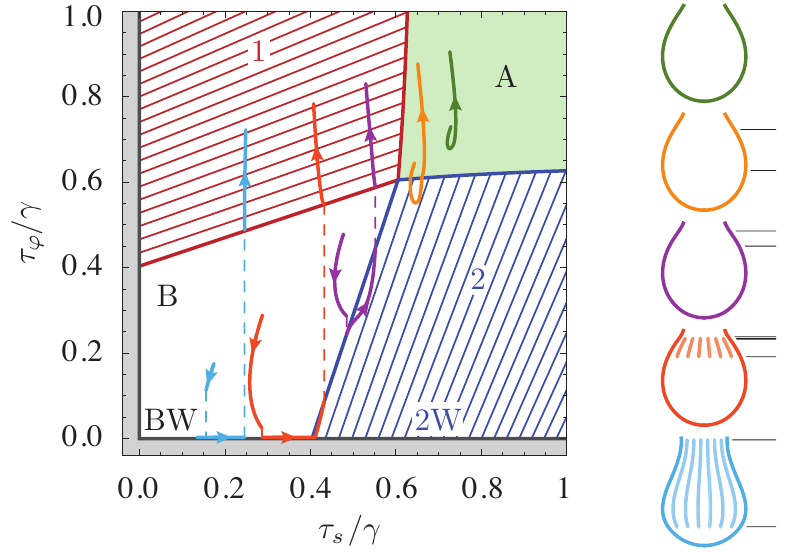}
 \caption {
  Trajectories in the stress plane for integrations at different
   stages of deflation (see pictograms on the right). Thin dashed parts of the
   trajectories are jumps in $\tau_\phi$. The integration always starts at the
   apex with $\tau_s = \tau_\phi$, that is, on the angle bisector, and runs
   through different domains of the stress plane. The parameters of the
   elastic model are $\EHo/\gamma = 5$, $\nu = 1/3$, and $L=0.95$. In the
   pictograms, thin horizontal lines indicate the transitions between the
   regions.}
\label{fig:trajectories}
\end{figure}

The modified shape equations are integrated from the apex $s_0 = 0$ to the
attachment point $s_0 = L_0$ to the capillary. On this way, the integration
will run through different domains and must switch to the appropriate shape
equations discussed in the previous section and Appendix \ref{app:II}. 
Figure~\ref{fig:trajectories}
shows typical trajectories of the integration in the stress plane, that is,
parametric plots of $(\tau_s(s_0), \tau_\phi(s_0))$ with $s_0 \in [0, L_0]$.

We name the different domains of the stress plane as follows: The admissible
domain (light green in Figure~\ref{fig:trajectories}) is abbreviated ``A'', the
red and blue ruled regions are termed ``1'' and ``2'' (because they stem from
boundary 1 and 2 in the stretch plane), and the white region is termed
``B''. Regions 2 and B also appear in wrinkled form, and 
we then call them ``2W'' and ``BW''.

In the numerical integration, an event handler must be introduced which
detects when the integration runs from one region into another. Changes from
region A to regions 1, 2 or B can be detected on the basis of strains, which
are limited by the boundary of the forbidden
domain as given by eq~\ref{eq:lambda_boundary}. The other direction, a
change from a hard-core region into the A domain, occurs when the hard-core
stresses become positive.

Switching between the different hard-core regions is, however, a bit more
complicated because the continuity conditions are less obvious. An elaborate
variational calculation \cite{KnocheDiss} shows surprisingly that $\tau_\phi$
may jump at transitions from B\rA BW, B\rA 2, 2\rA 1 and BW\rA 1; see also
Figure~\ref{fig:trajectories}.  
 The physical reason behind this behavior is
that $\tau_\phi$ is constitutively undetermined in region B: It may jump
without requiring the hoop stretch $\lambda_\phi$ to jump, which would be
unphysical and would lead to a ruptured shape. 
 This jump is necessary when
starting in region B, where we see in Appendix \ref{app:II}
 that the shooting parameter is
eliminated: The jump, or rather its arc-length coordinate $s_J$, serves as a
substitute shooting parameter. In the following discussion of each trajectory
shown in Figure~\ref{fig:trajectories}, this becomes more clear.

\begin{description}
\item[B\brA BW\brA 1, Light Blue.]  
 The transition from B to BW occurs at
  $s_J$ which can be chosen arbitrary (it just has to occur before the
  trajectory reaches the wrinkling region $\tau_\phi = 0$). This is the
  shooting parameter that is adjusted to match the boundary condition $r(L_0)
  = a/2$ at the end of integration. The rest of the trajectory follows 
  from the rules formulated above: 
   The jump from BW to 1 occurs when the wrinkling condition
  $\bar \lambda_\phi \leq \lambda_\phi$ becomes false. As the stretches are
  fixed to $L$ in B, BW and on the boundary of 1, the stretches are continuous
  at this transition, only the hoop stress jumps.

\item[B\brA BW\brA 2W\brA 2\brA 1, Light Red.]  
  The first transition occurs
  again at a free position $s_J$, and the remainder of the course 
   follows: BW\rA 2W is continuous and occurs when the ratio
  $\tauc_\phi/\tauc_s$ becomes larger than 3; 2W\rA 2 is also continuous and
  happens when $\bar \lambda_\phi$ becomes larger than $\lambda_\phi$; and
  2\rA 1, which has a jump in $\tau_\phi$, is determined by $\lambda_\phi$
  becoming larger than $L$.

\item[B\brA 2\brA 1, Violet.]  
  Again, the transition out of region B happens
  at a shooting parameter $s_J$, and the second transition follows as
  explained in the previous trajectory.

\item[A\brA 2\brA A, Dark Yellow.]  For this trajectory, the shooting
  parameter is $\tau_s(0)$ as usual, because we are starting in region A. The
  integration switches from A to 2 when the boundary
  in the stretch plane is reached
   according to eq \ref{eq:lambda_boundary}; and back to A
  when the hard-core stresses become positive. Both transitions are
  continuous. A transition from A to 1 is also possible and obeys the same
  reasoning.
\end{description}
The dark green trajectory is trivial because it stays in region A for the
whole integration. More paths are possible and have been worked out, but the
presented ones were most commonly met in the numerical investigations.

\section{Analysis of Computed Deflated Shapes}

With the newly developed shape equations, we can compute deflated shapes of
capsules according
 to the hard-core/soft-shell elasticity model. For the numerical
analysis, we use $\gamma$ as the tension unit and $a$ as the length
unit. Dimensionless quantities are denoted with a tilde. Specifically, the
shape equations depend on the reduced density difference $\tilde \rho = a^2
\Delta \rho g/\gamma$, the reduced pressure $\tilde p = ap/\gamma$ which
controls the capsule volume and the reduced surface Young modulus $\tEHo =
\EHo/\gamma$. Their dimensionless solutions contain the shape $(\tilde r,
\tilde z) = (r/a, z/a)$ and tensions $\tilde \tau_i = \tau_i/\gamma$ with $i
\in \{s, \phi\}$.

\begin{figure}
 \centering
 \includegraphics{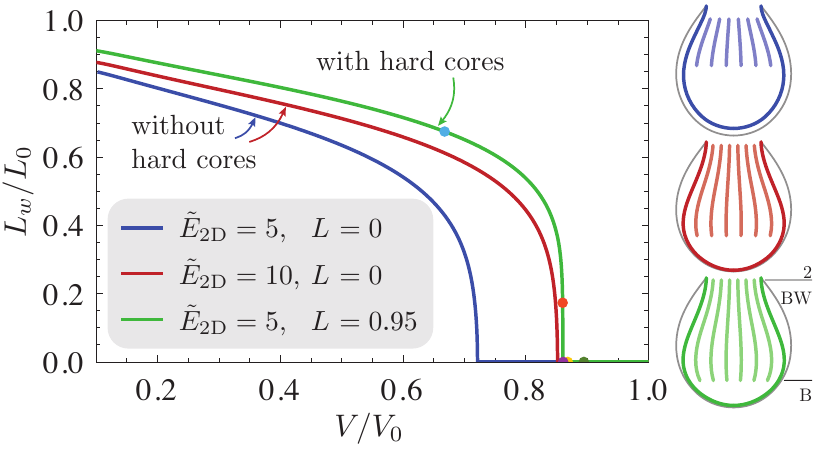}
 \caption { Wrinkle length as a function of the reduced volume for deflated
   shells with and without hard-core interactions, for $\nu = 1/3$ and $\tEHo$
   as indicated in the legend. The shapes on the right for $V = 0.7 V_0$
   reveal differences not only in $L_w$ but also in the overall shape (colors
   as in the legend, thin gray line is the undeformed shape, integration
   regions are indicated in the green contour).  The five data points on the
   green line in the diagram indicate the shapes and trajectories shown in
   Figure~\ref{fig:trajectories}.
}
\label{fig:hardcore_Lw}
\end{figure}

Starting with a Laplace-Young shape \cite{Knoche2013} with $\tilde \rho =
0.25$ and $\tilde p_0 = 2$, the pressure is lowered from $\tilde p = \tilde
p_0$ to $\tilde p = 0.1 \tilde p_0$. Figure~\ref{fig:hardcore_Lw} shows the
arc length $L_w$ of the wrinkled region as a function of the reduced volume
for three such series of deflated shapes with different elastic
parameters. The curves without hard-core interactions (red and blue, with
$L=0$) show that the onset of wrinkling occurs earlier for higher elastic
moduli. If hard-core interactions are included and occur already for small
compressions (green curve, $L = 0.95$), the wrinkling sets in early, even
though the elastic modulus is small. 
Here, the wrinkling is 
induced by hard cores coming into contact. After hard-core contact, 
 compressive 
stresses increase much more quickly with decreasing volume, 
as it can also  be seen in the 
larger ``stress loops'' in regions 1, 2, and B in the
 trajectories shown in Figure~\ref{fig:trajectories}. 
If these
stress loops touch the gray shaded wrinkling regions, wrinkling 
is triggered (red and blue shapes in Figure~\ref{fig:trajectories}). 
The quickly increasing stress loops after hard-core contact 
also give rise to the rather  steep increase of the 
arc length $L_w$ of the wrinkled region as a function of the reduced volume
in the corresponding green line in Figure~\ref{fig:hardcore_Lw}.

Thus, if one analyzes the shape of a deflated capsule, not knowing that it
obeys the hard-core/soft-shell elasticity model but assuming that it is a
usual Hookean membrane, the Hookean elastic modulus will be 
 overestimated considerably. 
Figure~\ref{fig:hardcore_Lw} illustrates this, as the green line
with $\tEHo = 5$ and $L= 0.95$ is much closer to the red line with $\tEHo =
10$ than to the blue line with $\tEHo = 5$.

\subsection{Shape Analysis of Theoretically Generated Shapes}

\begin{figure*}
 \centering
 \includegraphics{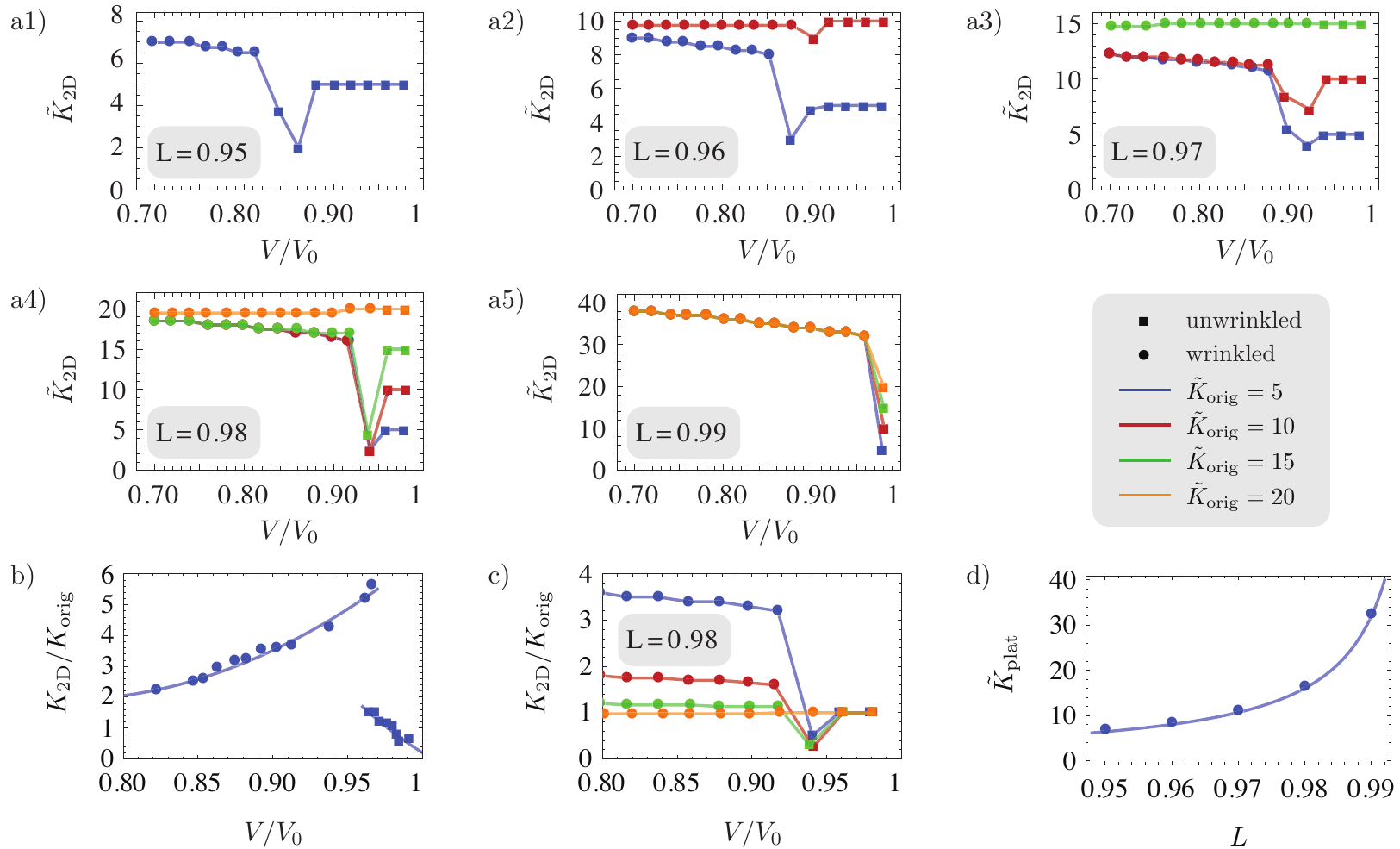}
 \caption{ 
  (a1--a5)
  Fit results for theoretically generated capsule shapes of various
   area compression moduli $\tilde K_\text{orig}$ of the soft shells 
   and hard-core lengths
   $L$. The fits are performed using the usual Hookean elasticity (without
   hard cores) and come to differing results $\tKHo$ for the fitted Hookean 
   area
   compression modulus. In both shape generation and fits, the Poisson ratio
   is fixed to $\nu = 1/3$. Shapes exhibiting 
   wrinkling are indicated by circles, unwrinkled shapes by squares.
   (b) 
 Typical  fit results for the  dimensionless 
   $\tKHo/\tilde K_\text{orig}$  area compression modulus of a
   hydrophobin capsule as obtained in ref~\citenum{Knoche2013}. 
The jump in the  elastic modulus coincides with the onset of wrinkling.
  (c) Fit results to theoretically 
   generated capsule shapes from subfigure (a4), normalized to $K_\text{orig}$.
   (d) The relation between the plateau value $\tilde K_\text{plat}$
  of  the fitted Hookean  modulus  $\tKHo$ and the hard-core length $L$ 
 according to  subfigures (a1)--(a5).
} 
\label{fig:hardcore_fit_results}
\end{figure*}

In ref~\citenum{Knoche2013}, a shape analysis for deflated pendant capsules was
developed. From experimental images, the contours are extracted and fitted
with the solutions of the shape equations (eqs \ref{eq:shape_pend}). This allows
one to determine the elastic modulus of the capsule membrane. The application of
this method to bubbles coated with a layer of the protein hydrophobin revealed
a very nonlinear elastic behavior, and the fitted elastic modulus as a
function of the volume jumps at the onset of wrinkling. The essential
features of the Hookean elastic modulus obtained in  ref~\citenum{Knoche2013}
are  shown in Figure~\ref{fig:hardcore_fit_results}b, where the 
area compression
modulus $\KHo = \EHo / 2(1-\nu)$ is used instead of the surface Young modulus.

We test whether this characteristic signature of the hydrophobin layer
elasticity can be explained by our hard-core/soft-shell elasticity model. 
To this
end, series of deflated shapes are computed using our modified shape
equations for a given area compression modulus $\tilde
K_\text{orig}$  for the soft springs and a given hard-core length $L$. 
Then   each theoretically generated shape is converted into a set of
sampling points and fed to the usual shape analysis algorithm developed in
ref~\citenum{Knoche2013} which uses a Hookean elasticity model without hard
cores. The output of the shape analysis algorithm is a 
fitted Hookean  area compression modulus $\tKHo$ for each 
capsule volume, which is shown in 
Figure~\ref{fig:hardcore_fit_results}a1-a5 
and demonstrates that the fitted Hookean 
 area  compression modulus $\tKHo$ can differ substantially 
from the value $\tilde K_\text{orig}$ of the soft springs 
that has been used for the theoretical shape generation and that 
the  fitted Hookean   area  compression modulus  can change with 
volume although   $\tilde K_\text{orig}$ and the hard-core length $L$
are fixed. 
In principle, the shape analysis algorithm using Hookean elasticity 
can also also give a fit value for the Poisson ratio $\nu$.
Alternatively, we can fix $\nu$ in the fit procedure to reduce the 
number of free parameters.

The analysis is concentrated to soft shell area compression moduli of $\tilde
K_\text{orig} = 5-20$ and hard-core lengths $L = 0.95-0.99$. 
Only for hard-core lengths $L>0.95$, we will reproduce the 
jump in the  fitted Hookean  area compression modulus $\tKHo$ 
in the volume range $V/V_0 > 0.95$, where it is also found for 
 bubbles coated with a layer of the protein hydrophobin 
in ref~\citenum{Knoche2013}. 
This motivates our choice of the  parameter range for $L$.
The range of  hard-core lengths  $L>0.95$ is higher than 
the above  estimates $L \approx 0.8-0.9$ from the 
results of refs~\citenum{Aumaitre2011} and \citenum{Aumaitre2012b} 
for planar monolayers.
These  differences can be caused by different 
surface densities of hydrophobin proteins 
in the bubble geometry as compared to the planar geometry:
In the preparation of bubbles \cite{Aumaitre2012, Knoche2013},
hydrophobins can adsorb from the surrounding 
bulk solution to the interface, which might lead to an increased 
surface density (and, thus, an increased $L$) due to the additional 
adsorption energy as compared to the planar 
geometry \cite{Aumaitre2011,Aumaitre2012b}, where a {\it fixed} amount 
of hydrophobin is spread onto the surface. 
In the fits using the  Hookean elasticity model, we 
fix the  Poisson ratio to the value $\nu = 1/3$, which 
is the appropriate value for a Hookean spring network, that is,
for small compression if hard cores do not come into contact.

When small hard-core lengths are combined with large area
compression moduli, for example, $L = 0.95$ 
with $\tilde K_\text{orig} = 20$, we 
find that the
hard-core interactions have no influence on the capsule 
shape, because the capsule starts to
wrinkle before the hard-cores come into contact. The numerical integration
then takes a path A\rA AW\rA A, and produces a shape that can also be produced
by purely  Hookean shape equations without hard-core interactions. 
Consequently, such  shapes can
be perfectly fitted with Hookean elasticity and the 
fitted Hookean  area compression modulus $\tKHo$ agrees with 
the correct  soft shell   area compression modulus $\tilde
K_\text{orig}$ as expected (see, for example, 
$L = 0.97$ with $\tilde K_\text{orig} = 15$ in
Figure~\ref{fig:hardcore_fit_results}a3).  
In particular, there is no jump in the fitted Hookean 
modulus $\tKHo$ during deflation.
Therefore, these cases are mostly omitted from
Figure~\ref{fig:hardcore_fit_results}a1--a5. 
There are some deflation series that are
very close to the limit where the hard-core interactions cease to influence
the shape (see $L=0.96$ and $\tilde K_\text{orig}=10$ in
Figure~\ref{fig:hardcore_fit_results}a2, for example), 
where there are only small
deviations between fitted  Hookean  and original soft shell modulus.

For computed deflation series  in Figure~\ref{fig:hardcore_fit_results}a1--a5
where the hard-core interactions profoundly influence the
shape, the fitted Hookean area compression modulus indeed shows 
 similar features as observed for hydrophobin capsules in
ref~\citenum{Knoche2013}  and shown  in 
Figure~\ref{fig:hardcore_fit_results}b.
For small deformations, where the hard cores are not yet in
contact, the fitted Hookean  modulus reproduces the original  soft shell 
value, $\tKHo = \tilde
K_\text{orig}$. When the hard cores come into contact, 
the fitted Hookean modulus $\tKHo$  exhibits a jump and 
can grow much larger than the original soft shell  value. 
For a better comparison between fit results for 
theoretical and experimental shapes, we 
show the dimensionless ratio $\tKHo/\tilde K_\text{orig}$ 
of fitted and soft shell area compression  moduli 
in Figure~\ref{fig:hardcore_fit_results}b and c (for the experimental fits in 
in Figure~\ref{fig:hardcore_fit_results}b 
we identify $\tilde K_\text{orig}$ with the 
mean of the fitted area compression modulus values before the jump). 
The peculiar dip in the fitted Hookean modulus 
$\tKHo$ just before the 
jump is an 
artifact of fixing  $\nu = 1/3$  in the fits. Further tests with free
$\nu$ show that these points can also be fitted with a Hookean 
modulus $\tKHo$ close to the original
soft shell  value $\tilde K_\text{orig}$ but with the fitted 
$\nu$ dropping to negative values; a result which 
lacks an intuitive explanation. 
Generally, the interpretation of the  fitted value for $\nu$
is not clear as it is obtained by using the  inappropriate 
Hookean elasticity model also in the regime, where hard cores come into 
contact.
The characteristic deflated volume, where  the 
fitted Hookean modulus $\tKHo$  exhibits a jump, increases 
with increasing hard-core length $L$ but 
depends only weakly on the  original
soft shell modulus $\tilde K_\text{orig}$.
 For hard-core lengths $L\ge 0.99$ 
close to unity, unwrinkled shapes are practically 
unobservable.

The deflation series  in Figure~\ref{fig:hardcore_fit_results}a1--a5 also 
shows that
contact of the hard cores triggers wrinkling (shapes exhibiting 
wrinkling are indicated by circles, unwrinkled shapes by squares). 
After the onset of wrinkling, the values for the fitted Hookean 
modulus  $\tKHo$ reach  a plateau
and increase only slightly for increasing deflation in the computed 
shapes in Figure~\ref{fig:hardcore_fit_results}a1-a5. 
For large $L>0.97$, 
the plateau value is
independent of the original soft shell modulus
 $\tilde K_\text{orig}$. 
The plateau value 
 strongly depends, however, 
 on the hard-core length $L$ as shown in
 Figure~\ref{fig:hardcore_fit_results}d. 
We can describe the observed plateau values 
by $\tilde K_\text{plat} \approx 0.3 (1-L)^{-1}$. 
This can be explained by assuming that both the fitted Hookean elasticity
and the hard-core/soft-shell elasticity have to describe the 
same sequence of shapes, the most prominent feature of which is the 
onset of wrinkles at a certain volume. 
If the onset of wrinkling is described by a fit with Hookean 
elasticity, the wrinkling criterion is  $\tau_\phi=0$, which gives
$\KHo (\lambda_\phi -1) \sim -\gamma$, see eq~\ref{eq:closing}
for the hoop stretch $\lambda_\phi$ at wrinkling. 
For hard-core/soft-shell elasticity wrinkling happens along the boundaries 
1 or 2 according to 
the criterion from eq \ref{eq:lambda_boundary}, which we approximate by
$\lambda_\phi \sim L$. If Hookean elasticity is to describe the 
 same shapes, $\lambda_\phi$ as a function of the volume 
has to be identical, in particular, at the onset of wrinkling, 
which results in $ 1-\lambda_\phi \sim  1/\tilde K_\text{plat}\sim  1-L$ 
or $\tilde K_\text{plat}\sim 1/(1-L)$.
Consequently, the fitted Hookean modulus  $\tKHo$ can exhibit a pronounced
jump  at the
onset of wrinkling if the original soft shell modulus 
 $\tilde K_\text{orig}$ is small and the hard-core length $L$ is large; in
Figure~\ref{fig:hardcore_fit_results}a5 
it jumps to more than its 6-fold value
for $\tilde K_\text{orig} = 5$ and $L = 0.99$.
The  hard cores have an influence on the elastic properties
if an increased plateau value is observed, that is, 
 $\tilde K_\text{plat} > \tilde K_\text{orig}$.
Therefore, the domain of influence of hard cores is given by 
$\tilde K_\text{orig} < 0.3 (1-L)^{-1}$, which means 
sufficiently large hard-core lengths.

We can use our findings for the  plateau value of the 
fitted Hookean modulues to extract the two 
hard-core/soft-shell elasticity parameters, the 
soft shell modulus   $\tilde K_\text{orig}= K_\text{orig}/\gamma$ and 
the hard-core length $L$, from the Hookean fits of deflated
shapes.
First, the fitted Hookean area compression modulus at small deflation 
before wrinkling can be used as an approximation to the soft shell modulus 
$K_\text{orig}$. 
For the hydrophbin capsule analyzed in ref~\citenum{Knoche2013} 
and shown in Figure~\ref{fig:hardcore_fit_results}b,
this gives a value of $K_\text{orig} = 342 \, \rm mN/m$.
Second, we can use  the above relation 
$\tilde K_\text{plat} \approx 0.3 (1-L)^{-1}$ to obtain the 
hard-core length $L$ from the plateau values of the fitted 
Hookean modulus.
The typical fit results for  hydrophobin capsules as 
 shown  in Figure~\ref{fig:hardcore_fit_results}b
do not show a plateau 
but  a gradual decrease of the fitted 
modulus after the pronounced jump for large deflations
and 
differ in this respect  from the fits of hard-core/soft-shell
capsules in Figure~\ref{fig:hardcore_fit_results}a.
Extracting hard-core lengths  $L \approx 1- 0.3/\tilde K_\text{plat}$,
from these fit results (using $\gamma = 49.8 {\rm mN/m}$), we find 
 values of the hard-core length $L$ which are decreasing
from $L=0.99$ at the jump to $L=0.98$.
This is a  hint that the $\beta$ barrel in hydrophobin 
is not an ideal hard core but weakly compressible.

Moreover,   hydrophobin capsules  also feature
an  initial increase of the 
fitted Hookean 
modulus for small  deflation (see Figure~\ref{fig:hardcore_fit_results}b)
which is absent  in all the fits of hard-core/soft-shell
capsules in
Figure~\ref{fig:hardcore_fit_results}a1--a5. 
This could reflect an additional nonlinear stiffening  
of the soft hydrophobin shells upon compression.

\section{Conclusions}

In this Article we developed an elasticity model for particle rafts at an
interface which consist of hard-core/soft-shell particles. Upon compression,
the membranes formed by these rafts first behave according to the Hookean
elastic law of the soft shells. 
Such ``soft shell behavior'' can be generated by any additional 
 interaction between hard-core particles, in principle. 
If the hard cores come into contact, further compression
is impeded. Additional stresses are then transmitted through the
``skeleton'' of hard cores, which must fulfill certain geometrical 
force balance  constraints; see eq~\ref{eq:tauc_ratios}.
This strongly modifies the elastic response, that is, the constitutive 
stress--strain  relations of the material, 
as soon as compression brings the hard cores of particles 
into contact. The model is characterized by an additional parameter 
$L<1$, which is the ratio of hard core diameter to the equilibrium lattice 
constant. 

For a planar particle layer under compression, we find 
that this rather general elastic model of a particle raft gives 
rise to a 
stress-dependent apparent Poisson ratio, which  increases upon compression 
beyond  $1/3$ and up to $\nu_\text{app}  = 1/(4L^2 - 1)$  
if hard cores come into contact; see Figure~\ref{fig:nu_app}.

Curved interfacial layers are important for the elasticity 
of particle decorated capsules. 
We reviewed the shape equations for capsules attached to a capillary and
modified them in order to include this hard-core/soft-shell elasticity model.
 This enabled us to compute deflated shapes. We find that the contact of 
hard cores by compression typically triggers wrinkling of the
capsule membrane because compressive 
stresses increase quickly with decreasing 
volume after hard-core contact, 
as illustrated in the stress diagram in Figure~\ref{fig:trajectories} 
and the steep increase of the wrinkle length for decreasing 
volume in Figure~\ref{fig:hardcore_Lw}.

In a further analysis, theoretically generated shapes
obeying the hard-core/soft-shell elasticity model were fitted with simple 
Hookean shape equations (without hard cores). 
The resulting fitted Hookean area
   compression modulus $\KHo$, 
as shown in Figure~\ref{fig:hardcore_fit_results}a1-a5
  for several parameter combinations of soft shell modulus and hard-core 
length,  exhibits features which are similar 
to the signatures obtained in ref~\citenum{Knoche2013} for 
hydrophobin capsules as shown in Figure~\ref{fig:hardcore_fit_results}b.
In particular we see a jump-like increase of the 
fitted Hookean modulus $\KHo$
at a characteristic deflated volume and a plateau value 
of  $\KHo$ for all smaller values. The jump in $\KHo$  
coincides with the volume where hard cores start to come into 
contact, which also causes the onset of wrinkling. 
This  explains the observed pronounced jump in the fitted 
Hookean area  compression modulus at the onset of wrinkling with 
the effect of the hard cores.
The characteristic deflated volume, where the jump in 
$\KHo$  occurs, increases with the  hard-core length $L$   approaching 
unity and is, thus, correlated 
 with an increasing plateau value  $\KHo$.
This suggests that the Hookean model is  only sufficient in a rather 
small volume range (above the jump in $\KHo$),
where shapes are   weakly deformed and  unwrinkled. 
To describe wrinkled shapes adequately, we need  to introduce 
the hard-core/soft-shell elasticity characterized by 
the  hard-core length $L$ and the Hookean modulus 
 $K_\text{orig}$ of the spring network. 
Nevertheless, the plateau value of $\KHo$  characterizes 
the effective  Hookean compression resistance of the particle 
raft in situations, 
where hard cores come into contact. 
We demonstrated that 
values of  the hard-core length $L$ and the Hookean modulus 
 $K_\text{orig}$ can already be inferred from a small set of distinct 
properties of the observed features of the fitted Hookean modulus
$\KHo$. For small deformations, before wrinkling and the associated 
jump of the modulus, we have $K_\text{orig}\approx \KHo$. The value of $L$
can be found from the plateau value of $\KHo$ after the jump.

The  hard-core/soft-shell model does not reproduce 
 the initial increase of the fitted Hookean 
compression modulus, nor its final
decrease as obtained in the hydrophobin fits.
The initial increase of the fitted 
 compression modulus might be reproducable if
nonlinear spring interactions are included in the elasticity model (springs
that  stiffen upon compression). The decrease after the jump, on
the other hand, might be considered as a decreasing hard-core length $L$,
because for smaller $L$, the plateau value is smaller. This could be modeled
by cores which can be slightly compressed. Replacing the hard-core
interactions by spring-interactions with relatively large spring constant
could achieve this.

In summary, with respect to modeling the elastic properties of hydrophobin 
coated interfaces, 
there are several indicators that the proposed elastic model is
too simple in distinguishing easily compressible domains and purely
incompressible ones. A softer transition might produce results closer to the
hydrophobin elasticity, modeled for example by springs whose spring constant
slightly increases with compression, and then sharply increases to a large but
finite hard-core spring constant. Such a model could even be theoretically 
more tractable because the resulting elastic law might involve a
bijective mapping between stretches and stresses if the spring constant is a
continuous function of the compression. 

The model is interesting not only with respect to hydrophobin 
coated liquid interfaces but also as a generic model for rafts of 
interacting hard particles at liquid interfaces, where the particle 
interactions give rise  to the soft shell elasticity.
Such type of membranes will exhibit a pronounced compression-stiffening 
after hard cores come into contact. For the capsules attached to 
capillaries this effect  induced wrinkling and led to a corresponding 
jump in the apparent or fitted Hookean are compression modulus.

The pronounced compression-stiffening of such membrane materials could 
also be used to stabilize structure, for example, {\em closed} 
capsules against compressive buckling \cite{Xu2005}.
The shape equations that we derived can be applied to the 
 buckling behavior of particle decorated 
liquid droplets in future work as well. 
We expect that the interacting particle raft will give rise 
to high resistance to buckling if it is engineered such that 
hard cores come into contact at the critical buckling pressure.

\begin{acknowledgements}
 We thank Dominic Vella for helpful remarks on the manuscript. 
 \end{acknowledgements}

\appendix

\section{
Force-Equilibrium of the Hard-Core stresses}
\label{app:tauc_ratios}

We consider a jammed state of the lattice, that is, when $(\lambda_x,
\lambda_y)$ is on the boundary of the admissible domain. Here, boundary 1 as
specified by eq~\ref{eq:lambda_boundary} and Figure~\ref{fig:ellipses}b is
considered, where $3 \lambda_x^2 / 4 + \lambda_y^2 / 4 = L^2$.

\begin{figure*}
 \centering 
 \includegraphics{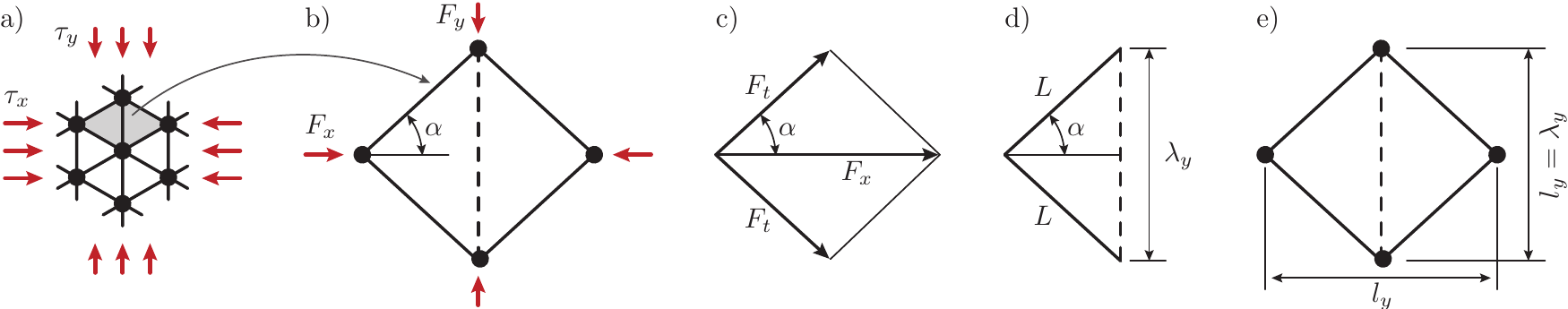}
 \caption{Calculation of the ratio of the hard-core stresses}
 \label{fig:force_ratio}
\end{figure*}

On this boundary, the lattice (see Figure~\ref{fig:force_ratio}a) is
compressed predominantly in the $x$-direction. 
Not all hard cores of neighboring
beads are in contact with each other, only those along the links that are
drawn continuous in Figure~\ref{fig:force_ratio}b. 
The dashed link in this figure is a spring
interaction, not a hard-core interaction, and is therefore ignored in the
following. Figure~\ref{fig:force_ratio}c 
shows that the force $F_x$ applied to a bead is split
into components $F_t$ tangential to the links. Trigonometric relations give
$\cos \alpha = F_x / 2 F_t$. Analogous considerations give $\sin \alpha = F_y
/ 2 F_t$ for the splitting of $F_y$. Thus, we have
\begin{equation}
 \frac{F_x}{F_y} = \tan \alpha
\end{equation} 
as the condition that the skeleton is in static equilibrium.

With Figure~\ref{fig:force_ratio}d, 
we can relate the geometric angle $\alpha$ to the lengths of
the links. The hard-core links have, by definition, length $L$. The vertical
(dashed) link has a rest length of $1$, and is stretched (or compressed) by
the deformation to a length $\lambda_y \cdot 1$. We thus obtain
\begin{equation}
 \tan \alpha = \frac{\lambda_y}{\sqrt{4 L^2 - \lambda_y^2}} 
  \quad \Rightarrow \quad
 \frac{F_x}{F_y} = \frac{\lambda_y}{\sqrt{4 L^2 - \lambda_y^2}}.
\end{equation} 
Finally, we have to relate the forces $F_x$ and $F_y$ to the stresses
$\tauc_x$ and $\tauc_y$. Stresses are forces per length, and the investigated
cell of the lattice has a height $l_y = \lambda_y$ and width $l_x = l_y/ \tan
\alpha = \sqrt{4 L^2 - \smash{\lambda_y^2}\vphantom{\lambda^2}}$; see
Figure~\ref{fig:force_ratio}e. 
With $\tauc_x = F_x/l_x$ and $\tauc_y = F_y/l_y$, we thus arrive
at
\begin{equation}
 \frac{\tauc_y}{\tauc_x} = \frac{\lambda_y^2}{4L^2 - \lambda_y^2}.
\end{equation} 
With the strain constraint from eq \ref{eq:lambda_boundary} on boundary 1, 
this is equivalent to
\begin{equation}
 \frac{\tauc_y}{\tauc_x} = \frac{1}{3} \frac{\lambda_y^2}{\lambda_x^2}.
\end{equation} 
Thus, only the ratio between the hard-core stresses is prescribed by the
geometry of the lattice.

Analogous results can be obtained on boundary 2, with all indices $x$ and $y$
interchanged. The complete result is therefore
\begin{equation} \label{eq:tauc_ratio_app}
 \frac{\tauc_y}{\tauc_x} = 
  \begin{cases}
   {\lambda_y^2}/3{\lambda_x^2} & \text{for } \lambda_y > L \\
   3 {\lambda_y^2}/{\lambda_x^2} & \text{for } \lambda_y < L
  \end{cases}.
\end{equation} 

In point B with $\lambda_x = \lambda_y = L$, which lies on both boundaries 1
and 2, the lattice is uniformly compressed and all neighbouring beads are in
contact. Equation (\ref{eq:tauc_ratio_app}) then states that the ratio of the
hard-core stresses is either $3$ or $1/3$. In fact, due to the close packing
of spheres, any value in between can also be realized, so that
\begin{equation} \label{eq:tauc_B_app}
 \frac{1}{3} \leq \frac{\tauc_y}{\tauc_x} \leq 3 \quad 
\text{for } \lambda_x = \lambda_y = L.
\end{equation} 
Equations (\ref{eq:tauc_ratio_app}) and (\ref{eq:tauc_B_app}) give
(\ref{eq:tauc_ratios}) in the main text.

\section{Hard-core/soft-shell elasticity  and 
   capsule shape equations}
\label{app:II}


We want to use the elasticity model for
particle rafts developed above (eqs \ref{eq:taus}, \ref{eq:stresses_complete}
 and \ref{eq:tauc_ratios}), including the hard-core interactions,
in the shape equations (eqs \ref{eq:shape_pend}) for capsules. 
Therefore, depending on the size of compressive stresses, 
we have to switch between different  constitutive elastic
relations corresponding to Hookean  constitutive relations 
(eqs \ref{eq:closing})
or hard-core  constitutive relations  
along the arc-length coordinate $s_0$. 

We
identify the $x$-direction of the planar model with the (meridional)
$s$-direction of the axisymmetric shell, and the $y$-direction with the
(circumferential) $\phi$-direction. With the help of the plots of the stretch
and stress planes in Figure~\ref{fig:schmetterlinge}c and d, we construct a 
suitable algorithm to calculate $\lambda_s$ and $\tau_\phi$ 
from given $\lambda_\phi$ and $\tau_s$: 
\begin{enumerate}
\item[(i)] We check if the point is in the admissible domain or
on the boundary by calculating $\lambdab_s(\lambda_\phi)$ with
 eq \ref{eq:lambda_boundary} and $\taus_s \big(\lambdab_s, \lambda_\phi \big)$
using eq \ref{eq:taus}, 
which is the smallest possible stress in the admissible
domain. If the given $\tau_s$ is larger (less compressive) 
than this value, the point is in the
admissible domain, if it is smaller (more compressive) 
then the point is on the boundary to the
forbidden domain. 
\item[(ii)] 
If the point is in the admissible domain,
eqs~\ref{eq:closing} can be used to calculate $\lambda_s$ and
$\tau_\phi$. 
\item[(iii)] 
If the point is on the boundary, then we know that
$\lambda_s = \lambdab_s(\lambda_\phi)$ according to
eq~\ref{eq:lambda_boundary}. In addition, we can calculate the spring
contributions $\taus_s\big(\lambdab_s, \lambda_\phi \big)$ from Hooke's law
(eq \ref{eq:taus}). The hard-core contribution can then be obtained as the
difference between the given $\tau_s$ and the spring contribution, $\tauc_s =
\tau_s - \taus_s$. This value should be negative. The sought stress
$\tau_\phi$ can then be calculated from the Hookean contribution $\taus_\phi$
according to eq~\ref{eq:taus} and the hard-core contribution $\tauc_\phi$
according to the force balance condition (eq \ref{eq:tauc_ratios}), 
so that $\tau_\phi = \taus_\phi \big(
\lambdab_s, \lambda_\phi \big) + \tauc_\phi$.
\end{enumerate}
Thus, we can use the above shape equations (eqs \ref{eq:shape_pend}) also 
for the
computation of deformed shapes for capsules obeying our hard-core/soft-shell
elasticity model. During the integration of the shape equations, however, 
the closing
relations (eqs \ref{eq:closing}) that are necessary to compute 
the right-hand side
must be replaced by the above procedure (iii) 
in regions where the lattice is on
boundary 1 or 2.

This method does not work when the lattice is stuck in point B, which must be
handled separately in the shape equations. The reason is that the confinement
to $\lambda_s = \lambda_\phi = L$ already determines the shape of the capsule:
It is uniformly compressed. The circumferential stretch $r/r_0 = L$ directly
implies $r(s_0) = L r_0(s_0)$. Inserting this solution into the differential
equation for $r$ in the system of shape equations (eqs \ref{eq:shape_pend}) 
then
yields $\psi(s_0) = \psi_0(s_0)$, where $\psi_0$ is the slope angle in the
undeformed configuration. The differential equation for $z$ then becomes
$z'(s_0) = L \sin \psi_0$ and has the solution $z(s_0) = L z_0(s_0) + c$ with
some constant $c$ depending on the starting value. From the differential
equation for $\psi$ in the shape equations (eqs \ref{eq:shape_pend}) we can then
deduce
\begin{equation} \label{eq:tau_phi_B}
 \tau_\phi = \left(- \frac{\kappa_{s_0}}{L} \tau_s + p 
   - \Delta\rho \, g \, z \right) / \kappa_\phi
\end{equation} 
by inserting the known solutions, where $\kappa_{s_0}$ is the meridional
curvature in the undeformed configuration. So in principle, only $\tau_s$ must
be determined by solving its differential equation. In order to keep the code
of the numerical implementation consistent, however, we solve the full system
of shape equations (eqs \ref{eq:shape_pend}) with the closing relations
(eqs \ref{eq:closing}) modified to contain the explicit result,
eq~\ref{eq:tau_phi_B}, just derived. This produces the correct numerical
solutions in the same ``data format'' as in all other parts.

When the lattice is in the jammed state B already at the start point of
integration (at the apex), we need to evaluate the limits of the explicit
solutions for $s_0 \rightarrow 0$. At the apex, the meridional and
circumferential curvatures coincide, $\kappa_s(0) = \kappa_\phi(0) =
\kappa_{s_0} / L = p_0 / 2 \gamma L$, where the last step can be derived from
the Laplace--Young equation and $p_0$ is the pressure inside the capsule in its
undeformed configuration.  From the force balance, the meridional and
circumferential tensions follow as $\tau_s(0) = \tau_\phi(0) = p L \gamma /
p_0$. This finding has a remarkable impact: $\tau_s(0)$ is fixed by the
external parameters, and cannot serve as a shooting parameter. The problem of
having lost the only shooting parameter is resolved in the main text 
 in the discussion of continuity conditions.

\bibliography{literature}

\begin{thebibliography}{43}%
\makeatletter
\providecommand \@ifxundefined [1]{%
 \@ifx{#1\undefined}
}%
\providecommand \@ifnum [1]{%
 \ifnum #1\expandafter \@firstoftwo
 \else \expandafter \@secondoftwo
 \fi
}%
\providecommand \@ifx [1]{%
 \ifx #1\expandafter \@firstoftwo
 \else \expandafter \@secondoftwo
 \fi
}%
\providecommand \natexlab [1]{#1}%
\providecommand \enquote  [1]{``#1''}%
\providecommand \bibnamefont  [1]{#1}%
\providecommand \bibfnamefont [1]{#1}%
\providecommand \citenamefont [1]{#1}%
\providecommand \href@noop [0]{\@secondoftwo}%
\providecommand \href [0]{\begingroup \@sanitize@url \@href}%
\providecommand \@href[1]{\@@startlink{#1}\@@href}%
\providecommand \@@href[1]{\endgroup#1\@@endlink}%
\providecommand \@sanitize@url [0]{\catcode `\\12\catcode `\$12\catcode
  `\&12\catcode `\#12\catcode `\^12\catcode `\_12\catcode `\%12\relax}%
\providecommand \@@startlink[1]{}%
\providecommand \@@endlink[0]{}%
\providecommand \url  [0]{\begingroup\@sanitize@url \@url }%
\providecommand \@url [1]{\endgroup\@href {#1}{\urlprefix }}%
\providecommand \urlprefix  [0]{URL }%
\providecommand \Eprint [0]{\href }%
\providecommand \doibase [0]{http://dx.doi.org/}%
\providecommand \selectlanguage [0]{\@gobble}%
\providecommand \bibinfo  [0]{\@secondoftwo}%
\providecommand \bibfield  [0]{\@secondoftwo}%
\providecommand \translation [1]{[#1]}%
\providecommand \BibitemOpen [0]{}%
\providecommand \bibitemStop [0]{}%
\providecommand \bibitemNoStop [0]{.\EOS\space}%
\providecommand \EOS [0]{\spacefactor3000\relax}%
\providecommand \BibitemShut  [1]{\csname bibitem#1\endcsname}%
\let\auto@bib@innerbib\@empty
\bibitem [{\citenamefont {Mooney}(1940)}]{Mooney1940}%
  \BibitemOpen
  \bibfield  {author} {\bibinfo {author} {\bibfnamefont {M.}~\bibnamefont
  {Mooney}},\ }\href {\doibase 10.1063/1.1712836} {\bibfield  {journal}
  {\bibinfo  {journal} {J. Appl. Phys.}\ }\textbf {\bibinfo {volume} {11}},\
  \bibinfo {pages} {582} (\bibinfo {year} {1940})}\BibitemShut {NoStop}%
\bibitem [{\citenamefont {Zhong-can}\ and\ \citenamefont
  {Helfrich}(1989)}]{Helfrich1989}%
  \BibitemOpen
  \bibfield  {author} {\bibinfo {author} {\bibfnamefont {O.-Y.}\ \bibnamefont
  {Zhong-can}}\ and\ \bibinfo {author} {\bibfnamefont {W.}~\bibnamefont
  {Helfrich}},\ }\href@noop {} {\bibfield  {journal} {\bibinfo  {journal}
  {Phys. Rev. A}\ }\textbf {\bibinfo {volume} {39}},\ \bibinfo {pages} {5280}
  (\bibinfo {year} {1989})}\BibitemShut {NoStop}%
\bibitem [{\citenamefont {Skalak}\ \emph {et~al.}(1973)\citenamefont {Skalak},
  \citenamefont {Tozeren}, \citenamefont {Zarda},\ and\ \citenamefont
  {Chien}}]{Skalak1973}%
  \BibitemOpen
  \bibfield  {author} {\bibinfo {author} {\bibfnamefont {R.}~\bibnamefont
  {Skalak}}, \bibinfo {author} {\bibfnamefont {A.}~\bibnamefont {Tozeren}},
  \bibinfo {author} {\bibfnamefont {R.}~\bibnamefont {Zarda}}, \ and\ \bibinfo
  {author} {\bibfnamefont {S.}~\bibnamefont {Chien}},\ }\href@noop {}
  {\bibfield  {journal} {\bibinfo  {journal} {Biophys. J.}\ }\textbf {\bibinfo
  {volume} {13}},\ \bibinfo {pages} {245} (\bibinfo {year} {1973})}\BibitemShut
  {NoStop}%
\bibitem [{\citenamefont {Evans}(1973{\natexlab{a}})}]{Evans1973a}%
  \BibitemOpen
  \bibfield  {author} {\bibinfo {author} {\bibfnamefont {E.}~\bibnamefont
  {Evans}},\ }\href@noop {} {\bibfield  {journal} {\bibinfo  {journal}
  {Biophys. J.}\ }\textbf {\bibinfo {volume} {13}},\ \bibinfo {pages} {926}
  (\bibinfo {year} {1973}{\natexlab{a}})}\BibitemShut {NoStop}%
\bibitem [{\citenamefont {Evans}(1973{\natexlab{b}})}]{Evans1973b}%
  \BibitemOpen
  \bibfield  {author} {\bibinfo {author} {\bibfnamefont {E.}~\bibnamefont
  {Evans}},\ }\href@noop {} {\bibfield  {journal} {\bibinfo  {journal}
  {Biophys. J.}\ }\textbf {\bibinfo {volume} {13}},\ \bibinfo {pages} {941}
  (\bibinfo {year} {1973}{\natexlab{b}})}\BibitemShut {NoStop}%
\bibitem [{\citenamefont {{Lim H W}}\ \emph {et~al.}(2002)\citenamefont {{Lim H
  W}}, \citenamefont {Wortis},\ and\ \citenamefont {Mukhopadhyay}}]{LimHW2002}%
  \BibitemOpen
  \bibfield  {author} {\bibinfo {author} {\bibfnamefont {G.}~\bibnamefont {{Lim
  H W}}}, \bibinfo {author} {\bibfnamefont {M.}~\bibnamefont {Wortis}}, \ and\
  \bibinfo {author} {\bibfnamefont {R.}~\bibnamefont {Mukhopadhyay}},\ }\href
  {\doibase 10.1073/pnas.202617299} {\bibfield  {journal} {\bibinfo  {journal}
  {Proc. Natl. Acad. Sci. U.S.A.}\ }\textbf {\bibinfo {volume} {99}},\ \bibinfo
  {pages} {16766} (\bibinfo {year} {2002})}\BibitemShut {NoStop}%
\bibitem [{\citenamefont {Lim}\ \emph {et~al.}(2008)\citenamefont {Lim},
  \citenamefont {Wortis},\ and\ \citenamefont {Mukhopadhy}}]{Lim2008}%
  \BibitemOpen
  \bibfield  {author} {\bibinfo {author} {\bibfnamefont {G.}~\bibnamefont
  {Lim}}, \bibinfo {author} {\bibfnamefont {M.}~\bibnamefont {Wortis}}, \ and\
  \bibinfo {author} {\bibfnamefont {R.}~\bibnamefont {Mukhopadhy}},\ }\href
  {\doibase 10.1002/9783527623372} {\emph {\bibinfo {title} {{Soft Matter}}}},\
  edited by\ \bibinfo {editor} {\bibfnamefont {G.}~\bibnamefont {Gompper}}\
  and\ \bibinfo {editor} {\bibfnamefont {M.}~\bibnamefont {Schick}}\ (\bibinfo
  {publisher} {Wiley-VCH Verlag GmbH \& Co. KGaA},\ \bibinfo {address}
  {Weinheim, Germany},\ \bibinfo {year} {2008})\BibitemShut {NoStop}%
\bibitem [{\citenamefont {Pickering}(1907)}]{Pickering1907}%
  \BibitemOpen
  \bibfield  {author} {\bibinfo {author} {\bibfnamefont {S.~U.}\ \bibnamefont
  {Pickering}},\ }\href {\doibase 10.1039/ct9079102001} {\bibfield  {journal}
  {\bibinfo  {journal} {J. Chem. Soc.}\ }\textbf {\bibinfo {volume} {91}},\
  \bibinfo {pages} {2001} (\bibinfo {year} {1907})}\BibitemShut {NoStop}%
\bibitem [{\citenamefont {Binks}(2002)}]{Binks2002}%
  \BibitemOpen
  \bibfield  {author} {\bibinfo {author} {\bibfnamefont {B.~P.}\ \bibnamefont
  {Binks}},\ }\href@noop {} {\bibfield  {journal} {\bibinfo  {journal} {Curr.
  Opin. Colloid Interface Sci.}\ }\textbf {\bibinfo {volume} {7}},\ \bibinfo
  {pages} {21} (\bibinfo {year} {2002})}\BibitemShut {NoStop}%
\bibitem [{\citenamefont {Horozov}(2008)}]{Horozov2008}%
  \BibitemOpen
  \bibfield  {author} {\bibinfo {author} {\bibfnamefont {T.}~\bibnamefont
  {Horozov}},\ }\href {\doibase 10.1016/j.cocis.2007.11.009} {\bibfield
  {journal} {\bibinfo  {journal} {Curr. Opin. Colloid Interface Sci.}\ }\textbf
  {\bibinfo {volume} {13}},\ \bibinfo {pages} {134} (\bibinfo {year}
  {2008})}\BibitemShut {NoStop}%
\bibitem [{\citenamefont {Dinsmore}\ \emph {et~al.}(2002)\citenamefont
  {Dinsmore}, \citenamefont {Hsu}, \citenamefont {Nikolaides}, \citenamefont
  {Marquez}, \citenamefont {Bausch},\ and\ \citenamefont
  {Weitz}}]{Dinsmore2002}%
  \BibitemOpen
  \bibfield  {author} {\bibinfo {author} {\bibfnamefont {A.~D.}\ \bibnamefont
  {Dinsmore}}, \bibinfo {author} {\bibfnamefont {M.~F.}\ \bibnamefont {Hsu}},
  \bibinfo {author} {\bibfnamefont {M.~G.}\ \bibnamefont {Nikolaides}},
  \bibinfo {author} {\bibfnamefont {M.}~\bibnamefont {Marquez}}, \bibinfo
  {author} {\bibfnamefont {A.~R.}\ \bibnamefont {Bausch}}, \ and\ \bibinfo
  {author} {\bibfnamefont {D.~A.}\ \bibnamefont {Weitz}},\ }\href {\doibase
  10.1126/science.1074868} {\bibfield  {journal} {\bibinfo  {journal}
  {Science}\ }\textbf {\bibinfo {volume} {298}},\ \bibinfo {pages} {1006}
  (\bibinfo {year} {2002})}\BibitemShut {NoStop}%
\bibitem [{\citenamefont {Bausch}\ \emph {et~al.}(2003)\citenamefont {Bausch},
  \citenamefont {Bowick}, \citenamefont {Cacciuto}, \citenamefont {Dinsmore},
  \citenamefont {Hsu}, \citenamefont {Nelson}, \citenamefont {Nikolaides},
  \citenamefont {Travesset},\ and\ \citenamefont {Weitz}}]{Bausch2003}%
  \BibitemOpen
  \bibfield  {author} {\bibinfo {author} {\bibfnamefont {A.~R.}\ \bibnamefont
  {Bausch}}, \bibinfo {author} {\bibfnamefont {M.~J.}\ \bibnamefont {Bowick}},
  \bibinfo {author} {\bibfnamefont {A.}~\bibnamefont {Cacciuto}}, \bibinfo
  {author} {\bibfnamefont {A.~D.}\ \bibnamefont {Dinsmore}}, \bibinfo {author}
  {\bibfnamefont {M.~F.}\ \bibnamefont {Hsu}}, \bibinfo {author} {\bibfnamefont
  {D.~R.}\ \bibnamefont {Nelson}}, \bibinfo {author} {\bibfnamefont {M.~G.}\
  \bibnamefont {Nikolaides}}, \bibinfo {author} {\bibfnamefont
  {A.}~\bibnamefont {Travesset}}, \ and\ \bibinfo {author} {\bibfnamefont
  {D.~A.}\ \bibnamefont {Weitz}},\ }\href {\doibase 10.1126/science.1081160}
  {\bibfield  {journal} {\bibinfo  {journal} {Science}\ }\textbf {\bibinfo
  {volume} {299}},\ \bibinfo {pages} {1716} (\bibinfo {year}
  {2003})}\BibitemShut {NoStop}%
\bibitem [{\citenamefont {Subramaniam}\ \emph {et~al.}(2005)\citenamefont
  {Subramaniam}, \citenamefont {Abkarian},\ and\ \citenamefont
  {Stone}}]{Subramaniam2005}%
  \BibitemOpen
  \bibfield  {author} {\bibinfo {author} {\bibfnamefont {A.~B.}\ \bibnamefont
  {Subramaniam}}, \bibinfo {author} {\bibfnamefont {M.}~\bibnamefont
  {Abkarian}}, \ and\ \bibinfo {author} {\bibfnamefont {H.}~\bibnamefont
  {Stone}},\ }\href {\doibase 10.1038/nmat1412} {\bibfield  {journal} {\bibinfo
   {journal} {Nat. Mater.}\ }\textbf {\bibinfo {volume} {4}},\ \bibinfo {pages}
  {553} (\bibinfo {year} {2005})}\BibitemShut {NoStop}%
\bibitem [{\citenamefont {Vella}\ \emph {et~al.}(2004)\citenamefont {Vella},
  \citenamefont {Aussillous},\ and\ \citenamefont {Mahadevan}}]{Vella2004}%
  \BibitemOpen
  \bibfield  {author} {\bibinfo {author} {\bibfnamefont {D.}~\bibnamefont
  {Vella}}, \bibinfo {author} {\bibfnamefont {P.}~\bibnamefont {Aussillous}}, \
  and\ \bibinfo {author} {\bibfnamefont {L.}~\bibnamefont {Mahadevan}},\ }\href
  {\doibase 10.1209/epl/i2004-10202-x} {\bibfield  {journal} {\bibinfo
  {journal} {Europhys. Lett.}\ }\textbf {\bibinfo {volume} {68}},\ \bibinfo
  {pages} {212} (\bibinfo {year} {2004})}\BibitemShut {NoStop}%
\bibitem [{\citenamefont {Zeng}\ \emph {et~al.}(2006)\citenamefont {Zeng},
  \citenamefont {Bissig},\ and\ \citenamefont {Dinsmore}}]{Zeng2006}%
  \BibitemOpen
  \bibfield  {author} {\bibinfo {author} {\bibfnamefont {C.}~\bibnamefont
  {Zeng}}, \bibinfo {author} {\bibfnamefont {H.}~\bibnamefont {Bissig}}, \ and\
  \bibinfo {author} {\bibfnamefont {A.}~\bibnamefont {Dinsmore}},\ }\href
  {\doibase 10.1016/j.ssc.2006.06.001} {\bibfield  {journal} {\bibinfo
  {journal} {Sol. State Commun.}\ }\textbf {\bibinfo {volume} {139}},\ \bibinfo
  {pages} {547} (\bibinfo {year} {2006})}\BibitemShut {NoStop}%
\bibitem [{\citenamefont {Cox}\ \emph {et~al.}(2007)\citenamefont {Cox},
  \citenamefont {Cagnol}, \citenamefont {Russell},\ and\ \citenamefont
  {Izzard}}]{Cox2007}%
  \BibitemOpen
  \bibfield  {author} {\bibinfo {author} {\bibfnamefont {A.~R.}\ \bibnamefont
  {Cox}}, \bibinfo {author} {\bibfnamefont {F.}~\bibnamefont {Cagnol}},
  \bibinfo {author} {\bibfnamefont {A.~B.}\ \bibnamefont {Russell}}, \ and\
  \bibinfo {author} {\bibfnamefont {M.~J.}\ \bibnamefont {Izzard}},\ }\href
  {\doibase 10.1021/la700451g} {\bibfield  {journal} {\bibinfo  {journal}
  {Langmuir}\ }\textbf {\bibinfo {volume} {23}},\ \bibinfo {pages} {7995}
  (\bibinfo {year} {2007})}\BibitemShut {NoStop}%
\bibitem [{\citenamefont {Stanimirova}\ \emph {et~al.}(2011)\citenamefont
  {Stanimirova}, \citenamefont {Marinova}, \citenamefont {Tcholakova},
  \citenamefont {Denkov}, \citenamefont {Stoyanov},\ and\ \citenamefont
  {Pelan}}]{Stanimirova2011}%
  \BibitemOpen
  \bibfield  {author} {\bibinfo {author} {\bibfnamefont {R.}~\bibnamefont
  {Stanimirova}}, \bibinfo {author} {\bibfnamefont {K.}~\bibnamefont
  {Marinova}}, \bibinfo {author} {\bibfnamefont {S.}~\bibnamefont
  {Tcholakova}}, \bibinfo {author} {\bibfnamefont {N.~D.}\ \bibnamefont
  {Denkov}}, \bibinfo {author} {\bibfnamefont {S.~D.}\ \bibnamefont
  {Stoyanov}}, \ and\ \bibinfo {author} {\bibfnamefont {E.}~\bibnamefont
  {Pelan}},\ }\href {\doibase 10.1021/la202860u} {\bibfield  {journal}
  {\bibinfo  {journal} {Langmuir}\ }\textbf {\bibinfo {volume} {27}},\ \bibinfo
  {pages} {12486} (\bibinfo {year} {2011})}\BibitemShut {NoStop}%
\bibitem [{\citenamefont {Ratanabanangkoon}\ \emph {et~al.}(2003)\citenamefont
  {Ratanabanangkoon}, \citenamefont {Gropper}, \citenamefont {Merkel},
  \citenamefont {Sackmann},\ and\ \citenamefont {Gast}}]{Ratanabanangkoon2003}%
  \BibitemOpen
  \bibfield  {author} {\bibinfo {author} {\bibfnamefont {P.}~\bibnamefont
  {Ratanabanangkoon}}, \bibinfo {author} {\bibfnamefont {M.}~\bibnamefont
  {Gropper}}, \bibinfo {author} {\bibfnamefont {R.}~\bibnamefont {Merkel}},
  \bibinfo {author} {\bibfnamefont {E.}~\bibnamefont {Sackmann}}, \ and\
  \bibinfo {author} {\bibfnamefont {A.~P.}\ \bibnamefont {Gast}},\ }\href
  {\doibase 10.1021/la026251y} {\bibfield  {journal} {\bibinfo  {journal}
  {Langmuir}\ }\textbf {\bibinfo {volume} {19}},\ \bibinfo {pages} {1054}
  (\bibinfo {year} {2003})}\BibitemShut {NoStop}%
\bibitem [{\citenamefont {W\"{o}sten}(2001)}]{Wosten2001}%
  \BibitemOpen
  \bibfield  {author} {\bibinfo {author} {\bibfnamefont {H.~A.}\ \bibnamefont
  {W\"{o}sten}},\ }\href {\doibase 10.1146/annurev.micro.55.1.625} {\bibfield
  {journal} {\bibinfo  {journal} {Annu. Rev. Microbiol.}\ }\textbf {\bibinfo
  {volume} {55}},\ \bibinfo {pages} {625} (\bibinfo {year} {2001})}\BibitemShut
  {NoStop}%
\bibitem [{\citenamefont {Hakanp\"{a}\"{a}}\ \emph {et~al.}(2004)\citenamefont
  {Hakanp\"{a}\"{a}}, \citenamefont {Paananen}, \citenamefont {Askolin},
  \citenamefont {Nakari-Set\"{a}l\"{a}}, \citenamefont {Parkkinen},
  \citenamefont {Penttil\"{a}}, \citenamefont {Linder},\ and\ \citenamefont
  {Rouvinen}}]{Hakanpaa2004}%
  \BibitemOpen
  \bibfield  {author} {\bibinfo {author} {\bibfnamefont {J.}~\bibnamefont
  {Hakanp\"{a}\"{a}}}, \bibinfo {author} {\bibfnamefont {A.}~\bibnamefont
  {Paananen}}, \bibinfo {author} {\bibfnamefont {S.}~\bibnamefont {Askolin}},
  \bibinfo {author} {\bibfnamefont {T.}~\bibnamefont {Nakari-Set\"{a}l\"{a}}},
  \bibinfo {author} {\bibfnamefont {T.}~\bibnamefont {Parkkinen}}, \bibinfo
  {author} {\bibfnamefont {M.}~\bibnamefont {Penttil\"{a}}}, \bibinfo {author}
  {\bibfnamefont {M.~B.}\ \bibnamefont {Linder}}, \ and\ \bibinfo {author}
  {\bibfnamefont {J.}~\bibnamefont {Rouvinen}},\ }\href {\doibase
  10.1074/jbc.M309650200} {\bibfield  {journal} {\bibinfo  {journal} {J. Biol.
  Chem.}\ }\textbf {\bibinfo {volume} {279}},\ \bibinfo {pages} {534} (\bibinfo
  {year} {2004})}\BibitemShut {NoStop}%
\bibitem [{\citenamefont {Linder}\ \emph {et~al.}(2005)\citenamefont {Linder},
  \citenamefont {Szilvay}, \citenamefont {Nakari-Set\"{a}l\"{a}},\ and\
  \citenamefont {Penttil\"{a}}}]{Linder2005}%
  \BibitemOpen
  \bibfield  {author} {\bibinfo {author} {\bibfnamefont {M.~B.}\ \bibnamefont
  {Linder}}, \bibinfo {author} {\bibfnamefont {G.~R.}\ \bibnamefont {Szilvay}},
  \bibinfo {author} {\bibfnamefont {T.}~\bibnamefont {Nakari-Set\"{a}l\"{a}}},
  \ and\ \bibinfo {author} {\bibfnamefont {M.~E.}\ \bibnamefont
  {Penttil\"{a}}},\ }\href {\doibase 10.1016/j.femsre.2005.01.004} {\bibfield
  {journal} {\bibinfo  {journal} {FEMS microbiology reviews}\ }\textbf
  {\bibinfo {volume} {29}},\ \bibinfo {pages} {877} (\bibinfo {year}
  {2005})}\BibitemShut {NoStop}%
\bibitem [{\citenamefont {Linder}(2009)}]{Linder2009}%
  \BibitemOpen
  \bibfield  {author} {\bibinfo {author} {\bibfnamefont {M.~B.}\ \bibnamefont
  {Linder}},\ }\href {\doibase 10.1016/j.cocis.2009.04.001} {\bibfield
  {journal} {\bibinfo  {journal} {Curr. Opin. Colloid Interface Sci.}\ }\textbf
  {\bibinfo {volume} {14}},\ \bibinfo {pages} {356} (\bibinfo {year}
  {2009})}\BibitemShut {NoStop}%
\bibitem [{\citenamefont {Basheva}\ \emph {et~al.}(2011)\citenamefont
  {Basheva}, \citenamefont {Kralchevsky}, \citenamefont {Christov},
  \citenamefont {Danov}, \citenamefont {Stoyanov}, \citenamefont
  {Blijdenstein}, \citenamefont {Kim}, \citenamefont {Pelan},\ and\
  \citenamefont {Lips}}]{Basheva2011}%
  \BibitemOpen
  \bibfield  {author} {\bibinfo {author} {\bibfnamefont {E.~S.}\ \bibnamefont
  {Basheva}}, \bibinfo {author} {\bibfnamefont {P.~A.}\ \bibnamefont
  {Kralchevsky}}, \bibinfo {author} {\bibfnamefont {N.~C.}\ \bibnamefont
  {Christov}}, \bibinfo {author} {\bibfnamefont {K.~D.}\ \bibnamefont {Danov}},
  \bibinfo {author} {\bibfnamefont {S.~D.}\ \bibnamefont {Stoyanov}}, \bibinfo
  {author} {\bibfnamefont {T.~B.~J.}\ \bibnamefont {Blijdenstein}}, \bibinfo
  {author} {\bibfnamefont {H.-J.}\ \bibnamefont {Kim}}, \bibinfo {author}
  {\bibfnamefont {E.~G.}\ \bibnamefont {Pelan}}, \ and\ \bibinfo {author}
  {\bibfnamefont {A.}~\bibnamefont {Lips}},\ }\href {\doibase
  10.1021/la104726w} {\bibfield  {journal} {\bibinfo  {journal} {Langmuir}\
  }\textbf {\bibinfo {volume} {27}},\ \bibinfo {pages} {2382} (\bibinfo {year}
  {2011})}\BibitemShut {NoStop}%
\bibitem [{\citenamefont {Gibbs}\ \emph {et~al.}(1999)\citenamefont {Gibbs},
  \citenamefont {Kermasha}, \citenamefont {Alli},\ and\ \citenamefont
  {Mulligan}}]{Gibbs1999}%
  \BibitemOpen
  \bibfield  {author} {\bibinfo {author} {\bibfnamefont {B.~F.}\ \bibnamefont
  {Gibbs}}, \bibinfo {author} {\bibfnamefont {S.}~\bibnamefont {Kermasha}},
  \bibinfo {author} {\bibfnamefont {I.}~\bibnamefont {Alli}}, \ and\ \bibinfo
  {author} {\bibfnamefont {C.~N.}\ \bibnamefont {Mulligan}},\ }\href {\doibase
  10.1080/096374899101256} {\bibfield  {journal} {\bibinfo  {journal} {Int. J.
  Food Sci. Nutr.}\ }\textbf {\bibinfo {volume} {50}},\ \bibinfo {pages} {213}
  (\bibinfo {year} {1999})}\BibitemShut {NoStop}%
\bibitem [{\citenamefont {Hektor}\ and\ \citenamefont
  {Scholtmeijer}(2005)}]{Hektor2005}%
  \BibitemOpen
  \bibfield  {author} {\bibinfo {author} {\bibfnamefont {H.~J.}\ \bibnamefont
  {Hektor}}\ and\ \bibinfo {author} {\bibfnamefont {K.}~\bibnamefont
  {Scholtmeijer}},\ }\href {\doibase 10.1016/j.copbio.2005.05.004} {\bibfield
  {journal} {\bibinfo  {journal} {Curr. Opin. Biotechnol.}\ }\textbf {\bibinfo
  {volume} {16}},\ \bibinfo {pages} {434} (\bibinfo {year} {2005})}\BibitemShut
  {NoStop}%
\bibitem [{\citenamefont {Lumsdon}\ \emph {et~al.}(2005)\citenamefont
  {Lumsdon}, \citenamefont {Green},\ and\ \citenamefont
  {Stieglitz}}]{Lumsdon2005}%
  \BibitemOpen
  \bibfield  {author} {\bibinfo {author} {\bibfnamefont {S.~O.}\ \bibnamefont
  {Lumsdon}}, \bibinfo {author} {\bibfnamefont {J.}~\bibnamefont {Green}}, \
  and\ \bibinfo {author} {\bibfnamefont {B.}~\bibnamefont {Stieglitz}},\ }\href
  {\doibase 10.1016/j.colsurfb.2005.06.012} {\bibfield  {journal} {\bibinfo
  {journal} {Colloids Surf. B Biointerfaces}\ }\textbf {\bibinfo {volume}
  {44}},\ \bibinfo {pages} {172} (\bibinfo {year} {2005})}\BibitemShut
  {NoStop}%
\bibitem [{\citenamefont {Tchuenbou-Magaia}\ \emph {et~al.}(2009)\citenamefont
  {Tchuenbou-Magaia}, \citenamefont {Norton},\ and\ \citenamefont
  {Cox}}]{Tchuenbou-Magaia2009}%
  \BibitemOpen
  \bibfield  {author} {\bibinfo {author} {\bibfnamefont {F.}~\bibnamefont
  {Tchuenbou-Magaia}}, \bibinfo {author} {\bibfnamefont {I.}~\bibnamefont
  {Norton}}, \ and\ \bibinfo {author} {\bibfnamefont {P.}~\bibnamefont {Cox}},\
  }\href {\doibase 10.1016/j.foodhyd.2009.03.005} {\bibfield  {journal}
  {\bibinfo  {journal} {Food Hydrocolloids}\ }\textbf {\bibinfo {volume}
  {23}},\ \bibinfo {pages} {1877} (\bibinfo {year} {2009})}\BibitemShut
  {NoStop}%
\bibitem [{\citenamefont {Knoche}\ \emph {et~al.}(2013)\citenamefont {Knoche},
  \citenamefont {Vella}, \citenamefont {Aumaitre}, \citenamefont {Degen},
  \citenamefont {Rehage}, \citenamefont {Cicuta},\ and\ \citenamefont
  {Kierfeld}}]{Knoche2013}%
  \BibitemOpen
  \bibfield  {author} {\bibinfo {author} {\bibfnamefont {S.}~\bibnamefont
  {Knoche}}, \bibinfo {author} {\bibfnamefont {D.}~\bibnamefont {Vella}},
  \bibinfo {author} {\bibfnamefont {E.}~\bibnamefont {Aumaitre}}, \bibinfo
  {author} {\bibfnamefont {P.}~\bibnamefont {Degen}}, \bibinfo {author}
  {\bibfnamefont {H.}~\bibnamefont {Rehage}}, \bibinfo {author} {\bibfnamefont
  {P.}~\bibnamefont {Cicuta}}, \ and\ \bibinfo {author} {\bibfnamefont
  {J.}~\bibnamefont {Kierfeld}},\ }\href {\doibase 10.1021/la402322g}
  {\bibfield  {journal} {\bibinfo  {journal} {Langmuir}\ }\textbf {\bibinfo
  {volume} {29}},\ \bibinfo {pages} {12463} (\bibinfo {year}
  {2013})}\BibitemShut {NoStop}%
\bibitem [{\citenamefont {Seung}\ and\ \citenamefont
  {Nelson}(1988)}]{Seung1988}%
  \BibitemOpen
  \bibfield  {author} {\bibinfo {author} {\bibfnamefont {H.~S.}\ \bibnamefont
  {Seung}}\ and\ \bibinfo {author} {\bibfnamefont {D.~R.}\ \bibnamefont
  {Nelson}},\ }\href@noop {} {\bibfield  {journal} {\bibinfo  {journal} {Phys.
  Rev. A}\ }\textbf {\bibinfo {volume} {38}},\ \bibinfo {pages} {1005}
  (\bibinfo {year} {1988})}\BibitemShut {NoStop}%
\bibitem [{\citenamefont {Landau}\ and\ \citenamefont
  {Lifshitz}(1987)}]{Landau1987}%
  \BibitemOpen
  \bibfield  {author} {\bibinfo {author} {\bibfnamefont {L.}~\bibnamefont
  {Landau}}\ and\ \bibinfo {author} {\bibfnamefont {E.}~\bibnamefont
  {Lifshitz}},\ }\href@noop {} {\emph {\bibinfo {title} {Fluid
  {M}echanics.}}},\ \bibinfo {edition} {2nd}\ ed.,\ Course of Theoretical
  Physics\ (\bibinfo  {publisher} {Butterworth-Heinemann},\ \bibinfo {address}
  {Oxford},\ \bibinfo {year} {1987})\BibitemShut {NoStop}%
\bibitem [{\citenamefont {Ostoja-Starzewski}(2002)}]{Ostoja-Starzewski2002}%
  \BibitemOpen
  \bibfield  {author} {\bibinfo {author} {\bibfnamefont {M.}~\bibnamefont
  {Ostoja-Starzewski}},\ }\href {\doibase 10.1115/1.1432990} {\bibfield
  {journal} {\bibinfo  {journal} {Appl. Mech. Rev.}\ }\textbf {\bibinfo
  {volume} {55}},\ \bibinfo {pages} {35} (\bibinfo {year} {2002})}\BibitemShut
  {NoStop}%
\bibitem [{\citenamefont {Libai}\ and\ \citenamefont
  {Simmonds}(1998)}]{Libai1998}%
  \BibitemOpen
  \bibfield  {author} {\bibinfo {author} {\bibfnamefont {A.}~\bibnamefont
  {Libai}}\ and\ \bibinfo {author} {\bibfnamefont {J.~G.}\ \bibnamefont
  {Simmonds}},\ }\href@noop {} {\emph {\bibinfo {title} {{The Nonlinear Theory
  of Elastic Shells}}}}\ (\bibinfo  {publisher} {Cambridge University Press},\
  \bibinfo {year} {1998})\BibitemShut {NoStop}%
\bibitem [{\citenamefont {Cicuta}\ and\ \citenamefont
  {Vella}(2009)}]{Cicuta2009}%
  \BibitemOpen
  \bibfield  {author} {\bibinfo {author} {\bibfnamefont {P.}~\bibnamefont
  {Cicuta}}\ and\ \bibinfo {author} {\bibfnamefont {D.}~\bibnamefont {Vella}},\
  }\href {\doibase 10.1103/PhysRevLett.102.138302} {\bibfield  {journal}
  {\bibinfo  {journal} {Phys. Rev. Lett.}\ }\textbf {\bibinfo {volume} {102}},\
  \bibinfo {pages} {138302} (\bibinfo {year} {2009})}\BibitemShut {NoStop}%
\bibitem [{\citenamefont {Aumaitre}\ \emph {et~al.}(2011)\citenamefont
  {Aumaitre}, \citenamefont {Vella},\ and\ \citenamefont
  {Cicuta}}]{Aumaitre2011}%
  \BibitemOpen
  \bibfield  {author} {\bibinfo {author} {\bibfnamefont {E.}~\bibnamefont
  {Aumaitre}}, \bibinfo {author} {\bibfnamefont {D.}~\bibnamefont {Vella}}, \
  and\ \bibinfo {author} {\bibfnamefont {P.}~\bibnamefont {Cicuta}},\ }\href
  {\doibase 10.1039/c0sm01213k} {\bibfield  {journal} {\bibinfo  {journal}
  {Soft Matter}\ }\textbf {\bibinfo {volume} {7}},\ \bibinfo {pages} {2530}
  (\bibinfo {year} {2011})}\BibitemShut {NoStop}%
\bibitem [{\citenamefont {Landau}\ and\ \citenamefont
  {Lifshitz}(1970)}]{Landau1970}%
  \BibitemOpen
  \bibfield  {author} {\bibinfo {author} {\bibfnamefont {L.}~\bibnamefont
  {Landau}}\ and\ \bibinfo {author} {\bibfnamefont {E.}~\bibnamefont
  {Lifshitz}},\ }\href@noop {} {\emph {\bibinfo {title} {Theory of
  {E}lasticity.}}},\ \bibinfo {edition} {2nd}\ ed.,\ Course of Theoretical
  Physics\ (\bibinfo  {publisher} {Pergamon Press},\ \bibinfo {address}
  {Oxford},\ \bibinfo {year} {1970})\BibitemShut {NoStop}%
\bibitem [{\citenamefont {Clegg}(2008)}]{Clegg2008}%
  \BibitemOpen
  \bibfield  {author} {\bibinfo {author} {\bibfnamefont {P.~S.}\ \bibnamefont
  {Clegg}},\ }\href {\doibase 10.1088/0953-8984/20/11/113101} {\bibfield
  {journal} {\bibinfo  {journal} {J. Phys.: Condens. Matter}\ }\textbf
  {\bibinfo {volume} {20}},\ \bibinfo {pages} {113101} (\bibinfo {year}
  {2008})}\BibitemShut {NoStop}%
\bibitem [{\citenamefont {Aumaitre}(2012)}]{Aumaitre2012}%
  \BibitemOpen
  \bibfield  {author} {\bibinfo {author} {\bibfnamefont {E.}~\bibnamefont
  {Aumaitre}},\ }\emph {\bibinfo {title} {Viscoelastic Properties of
  {H}ydrophobin Layers}},\ \href@noop {} {Ph.D. thesis},\ \bibinfo  {school}
  {University of Cambridge} (\bibinfo {year} {2012})\BibitemShut {NoStop}%
\bibitem [{\citenamefont {Kisko}\ \emph {et~al.}(2009)\citenamefont {Kisko},
  \citenamefont {Szilvay}, \citenamefont {Vuorimaa}, \citenamefont
  {Lemmetyinen}, \citenamefont {Linder}, \citenamefont {Torkkeli},\ and\
  \citenamefont {Serimaa}}]{Kisko2009}%
  \BibitemOpen
  \bibfield  {author} {\bibinfo {author} {\bibfnamefont {K.}~\bibnamefont
  {Kisko}}, \bibinfo {author} {\bibfnamefont {G.~R.}\ \bibnamefont {Szilvay}},
  \bibinfo {author} {\bibfnamefont {E.}~\bibnamefont {Vuorimaa}}, \bibinfo
  {author} {\bibfnamefont {H.}~\bibnamefont {Lemmetyinen}}, \bibinfo {author}
  {\bibfnamefont {M.~B.}\ \bibnamefont {Linder}}, \bibinfo {author}
  {\bibfnamefont {M.}~\bibnamefont {Torkkeli}}, \ and\ \bibinfo {author}
  {\bibfnamefont {R.}~\bibnamefont {Serimaa}},\ }\href {\doibase
  10.1021/la803252g} {\bibfield  {journal} {\bibinfo  {journal} {Langmuir}\
  }\textbf {\bibinfo {volume} {25}},\ \bibinfo {pages} {1612} (\bibinfo {year}
  {2009})}\BibitemShut {NoStop}%
\bibitem [{\citenamefont {Aumaitre}\ \emph {et~al.}(2012)\citenamefont
  {Aumaitre}, \citenamefont {Wongsuwarn}, \citenamefont {Rossetti},
  \citenamefont {Hedges}, \citenamefont {Cox}, \citenamefont {Vella},\ and\
  \citenamefont {Cicuta}}]{Aumaitre2012b}%
  \BibitemOpen
  \bibfield  {author} {\bibinfo {author} {\bibfnamefont {E.}~\bibnamefont
  {Aumaitre}}, \bibinfo {author} {\bibfnamefont {S.}~\bibnamefont
  {Wongsuwarn}}, \bibinfo {author} {\bibfnamefont {D.}~\bibnamefont
  {Rossetti}}, \bibinfo {author} {\bibfnamefont {N.~D.}\ \bibnamefont
  {Hedges}}, \bibinfo {author} {\bibfnamefont {A.~R.}\ \bibnamefont {Cox}},
  \bibinfo {author} {\bibfnamefont {D.}~\bibnamefont {Vella}}, \ and\ \bibinfo
  {author} {\bibfnamefont {P.}~\bibnamefont {Cicuta}},\ }\href {\doibase
  10.1039/c1sm06139a} {\bibfield  {journal} {\bibinfo  {journal} {Soft Matter}\
  }\textbf {\bibinfo {volume} {8}},\ \bibinfo {pages} {1175} (\bibinfo {year}
  {2012})}\BibitemShut {NoStop}%
\bibitem [{\citenamefont {Steigmann}(1990)}]{Steigmann1990}%
  \BibitemOpen
  \bibfield  {author} {\bibinfo {author} {\bibfnamefont {D.~J.}\ \bibnamefont
  {Steigmann}},\ }\href {http://www.jstor.org/stable/51778} {\bibfield
  {journal} {\bibinfo  {journal} {Proc. R. Soc. London. A}\ }\textbf {\bibinfo
  {volume} {429}},\ \bibinfo {pages} {pp. 141} (\bibinfo {year}
  {1990})}\BibitemShut {NoStop}%
\bibitem [{\citenamefont {Davidovitch}\ \emph {et~al.}(2011)\citenamefont
  {Davidovitch}, \citenamefont {Schroll}, \citenamefont {Vella}, \citenamefont
  {Adda-Bedia},\ and\ \citenamefont {Cerda}}]{Davidovitch2011}%
  \BibitemOpen
  \bibfield  {author} {\bibinfo {author} {\bibfnamefont {B.}~\bibnamefont
  {Davidovitch}}, \bibinfo {author} {\bibfnamefont {R.~D.}\ \bibnamefont
  {Schroll}}, \bibinfo {author} {\bibfnamefont {D.}~\bibnamefont {Vella}},
  \bibinfo {author} {\bibfnamefont {M.}~\bibnamefont {Adda-Bedia}}, \ and\
  \bibinfo {author} {\bibfnamefont {E.}~\bibnamefont {Cerda}},\ }\href
  {\doibase 10.1073/pnas.1108553108} {\bibfield  {journal} {\bibinfo  {journal}
  {Proc. Natl. Acad. Sci. U.S.A.}\ }\textbf {\bibinfo {volume} {108}},\
  \bibinfo {pages} {18227} (\bibinfo {year} {2011})}\BibitemShut {NoStop}%
\bibitem [{\citenamefont {Knoche}(2014)}]{KnocheDiss}%
  \BibitemOpen
  \bibfield  {author} {\bibinfo {author} {\bibfnamefont {S.}~\bibnamefont
  {Knoche}},\ }\emph {\bibinfo {title} {{Instabilities and Shape Analyses of
  Elastic Shells}}},\ \href@noop {} {\bibinfo {type} {Phd thesis}},\ \bibinfo
  {school} {TU Dortmund} (\bibinfo {year} {2014})\BibitemShut {NoStop}%
\bibitem [{\citenamefont {Xu}\ \emph {et~al.}(2005)\citenamefont {Xu},
  \citenamefont {Melle}, \citenamefont {Golemanov},\ and\ \citenamefont
  {Fuller}}]{Xu2005}%
  \BibitemOpen
  \bibfield  {author} {\bibinfo {author} {\bibfnamefont {H.}~\bibnamefont
  {Xu}}, \bibinfo {author} {\bibfnamefont {S.}~\bibnamefont {Melle}}, \bibinfo
  {author} {\bibfnamefont {K.}~\bibnamefont {Golemanov}}, \ and\ \bibinfo
  {author} {\bibfnamefont {G.}~\bibnamefont {Fuller}},\ }\href {\doibase
  10.1021/la0507378} {\bibfield  {journal} {\bibinfo  {journal} {Langmuir}\
  }\textbf {\bibinfo {volume} {21}},\ \bibinfo {pages} {10016} (\bibinfo {year}
  {2005})}\BibitemShut {NoStop}%
\end{thebibliography}%

\end{document}